\documentclass[a4paper,12pt]{article}
\usepackage{jheppub}
\usepackage[dvipsnames]{xcolor}
\usepackage{tikz}
\usepackage{pgfplots}
\usepackage[T1]{fontenc}
\usepackage[]{slashed}
\usepackage[]{bm}     
\usepackage{physics}
\usepackage{dsfont}

\usepackage[force]{feynmp-auto}

\usepackage{hyperref}
\hypersetup{colorlinks=true,linkcolor=magenta,anchorcolor=green,citecolor=cyan,filecolor=black,menucolor=black,urlcolor=brown}
\makeatletter

\usepackage{lipsum}

\DeclareMathOperator\arccosh{arccosh}

\DeclareMathOperator\Li{Li}

\usepackage{graphicx}
\usepackage{epstopdf}
\usepackage{bbm,amsmath,graphicx,amssymb,amsfonts,amsthm}

\def\rddots#1{\cdot^{\cdot^{\cdot^{#1}}}}

\usepackage{soul}
\definecolor{bubbles}{rgb}{0.91, 1.0, 1.0}
\definecolor{aquamarine}{rgb}{0.5, 1.0, 0.83}
\definecolor{bubblegum}{rgb}{0.99, 0.76, 0.8}
\definecolor{blackbell}{rgb}{0.64, 0.64, 0.82}
\definecolor{dollarbill}{rgb}{0.72, 0.93, 0.6}

\def\sq[#1,#2]{\left[#1\,#2\right]}
\def\an[#1,#2]{\left\langle#1\,#2\right\rangle}

\def\an[#1,#2]{\left\langle#1\,#2\right\rangle}
\def\aq[#1,#2,#3]{\left\langle#1|#2|#3\right]}
\def\qa[#1,#2,#3]{\left[#1|#2|#3\right\rangle}
\def\sq[#1,#2]{\left[#1\,#2\right]}
\def\spa#1.#2{\left\langle#1\,#2\right\rangle}
\def\spab[#1,#2,#3]{\left\langle#1|#2|#3\right]}
\def\spba[#1,#2,#3]{\left[#1|#2|#3\right\rangle}
\def\spb#1.#2{\left[#1\,#2\right]}

\def\Ttrma(#1,#2,#3,#4){{\rm tr}_{-}[\slash \!\!\!\;\!\! #1\slash  \!\!\!\;\!\! #2 \slash  \!\!\!\;\!\!#3\slash  \!\!\!\;\!\!#4]}
\def\Ttrmb(#1,#2,#3,#4,#5,#6){{\rm tr}_{-}[\slash \!\!\!\;\!\! #1\slash  \!\!\!\;\!\! #2 \slash  \!\!\!\;\!\!#3\slash  \!\!\!\;\!\!#4\slash  \!\!\!\;\!\!#5\slash  \!\!\!\;\!\!#6]}
\def\Ttrmc(#1,#2,#3,#4,#5,#6,#7,#8){{\rm tr}_{-}[\slash \!\!\!\;\!\! #1\slash  \!\!\!\;\!\! #2 \slash  \!\!\!\;\!\!#3\slash  \!\!\!\;\!\!#4\slash  
\!\!\!\;\!\!#5\slash  \!\!\!\;\!\!#6\slash  \!\!\!\;\!\!#7\slash  \!\!\!\;\!\!#8]}
\def\Dp(#1,#2){(#1\cdot #2)}

\def\triangleboxleft{\scalebox{.9}{$\triangleleft$}\kern-.1em\Box}
\def\triangleboxright{\Box\kern-.1em\scalebox{.9}{$\triangleright$}}
\def\dBox{\Box\kern-.1em\Box}
\def\dNPBoxs{\scalebox{.9}{$\bowtie$}\kern-.1em\Box}
\def\dNPBoxu{\Box\kern-.1em\scalebox{.9}{$\bowtie$}}

\def\beq{\begin{equation}}
\def\eeq{\end{equation}}
\def\bes{\begin{split}}
\def\ees{\end{split}}
\def\beqa{\begin{eqnarray}}
\def\eeqa{\end{eqnarray}}

\def\eeqa{\end{eqnarray}}
\def\ek[#1,#2]{(\varepsilon_{#1}\cdot k_{#2})}
\def\e[#1,#2]{(\varepsilon_{#1}\cdot \varepsilon_{#2})}
\def\s(#1,#2){{(\ell_#1\cdot\ell_#2)}}

\def\e{\epsilon}

\pgfdeclarelayer{bg}    
\pgfsetlayers{bg,main}  

\definecolor{Mathematica}{HTML}{ed192d}

\usepackage{float}

\usetikzlibrary{shapes.misc}
\usetikzlibrary{arrows}
\usetikzlibrary{decorations.markings}
\usetikzlibrary{positioning,fit}
\usetikzlibrary{patterns}
\usetikzlibrary{decorations.pathmorphing}

\tikzset{snake it/.style={decorate, decoration=snake}}

\tikzset{cross/.style={cross out, draw=black, minimum size=2*(#1-\pgflinewidth), inner sep=0pt, outer sep=0pt},
	cross/.default={2pt}}

 \preprint{CERN-TH-2023-135, IPhT-T23/041,  LAPTh-029/23}
      
\title{Classical Observables from the Exponential Representation of the Gravitational S-Matrix }

\author[a,b]{\!\! Poul H. Damgaard}
\author[a]{\!\!, Elias Roos Hansen}
\author[a]{\!\!, Ludovic Plant\'e}
\author[c]{\!\!, Pierre Vanhove}

\affiliation[a]{Niels Bohr International Academy, Niels Bohr Institute, University of Copenhagen, Blegdamsvej 17, DK-2100 Copenhagen, Denmark}
\affiliation[b]{Theoretical Physics Department, CERN, 1211 Geneva 23, Switzerland}
\affiliation[c]{Institut de Physique Theorique, Universit\'e Paris-Saclay,
CEA, CNRS, F-91191 Gif-sur-Yvette Cedex, France}
\keywords{Scattering Amplitudes, General Relativity}

\abstract{By combining the KMOC-formalism with the exponential representation of
the scattering matrix we show that the two-body scattering angle is given by the corresponding matrix element of
the exponential representation. This holds to all
orders in the Post-Minkowskian expansion of gravity when restricted to the conservative sector. Once gravitational
radiation is taken into account new terms correcting this relationship appear starting at fourth Post-Minkowskian order. A systematic
expansion of the momentum kick is provided to any order, thus illustrating the iterative structure that partly recycles terms from 
lower orders in the Post-Minkowskian expansion. We provide explicit results for this computation to fourth
Post-Minkowskian order, the first complete calculation at this order based on scattering amplitudes.}

\begin{document} 
\maketitle
\flushbottom
\newpage
\section{Introduction}\label{sec:intro}

While the Post-Minkowskian expansion of general relativity~\cite{Damour:2016gwp,Damour:2017zjx,Bjerrum-Bohr:2018xdl,Cheung:2018wkq,Damour:2019lcq} has been highly 
successful in solving the relativistic two-body problem by means of modern amplitude techniques, new and puzzling features seem to appear at every new order
considered. The second-order Post-Minkowskian solution of Westpfahl~\cite{Westpfahl:1985tsl} was easily reproduced by amplitude methods \cite{Bjerrum-Bohr:2018xdl} but
already the first solution to third Post-Minkowskian order \cite{Bern:2019nnu,Bern:2019crd} displayed an unphysical divergence in the scattering angle
that could not be understood within the conservative framework used. The resolution was to be found when including radiation reaction of the gravitational
field~\cite{DiVecchia:2020ymx,Damour:2020tta,Bini:2021gat,DiVecchia:2021ndb,DiVecchia:2021bdo}. Remarkably, soft gravitons cancelled
the unwanted divergence in the scattering angle, thereby reproducing the classic result of Amati, Ciafaloni, and Veneziano \cite{Amati:1993tb}. Moreover, 
to this third Post-Minkowskian order a standard quantum field theoretic evaluation of
the full classical part of the gravitational two-to-two scattering amplitude precisely yields the correct scattering angle \cite{Bjerrum-Bohr:2021vuf,Bjerrum-Bohr:2021din},
the simple resolution being found in the need to include {\em all} classical pieces from the two-loop scattering amplitude. As explained in the latter two references,
those classical parts can be systematically identified through the so-called velocity cuts of the scattering amplitude: delta-function contributions that emerge from
combinations of propagators with the Feynman $i\epsilon$-prescription. For reviews of these ideas see, $e.g.$, ref. \cite{Bjerrum-Bohr:2022blt,Bjerrum-Bohr:2022ows}.

Among the many lessons learned at that third Post-Minkowskian order has been the need to understand how to subtract terms that diverge in the classical limit
in order to yield unambiguously those parts of the scattering amplitude that remain finite when $\hbar \to 0$. These delicate cancellations have their root in the
conventional use of the Born expansion of quantum field theory. Parametrizing the $S$-matrix as $\hat{S} = 1 + i\hat{T}/\hbar$, unitarity of  $\hat{S}$ leads to the
optical theorem through
\beq
\hat{T} - \hat{T}^{\dagger} ~=~ \frac{i}{\hbar}\hat{T}\hat{T}^{\dagger} ~.
\label{Tunit}
\eeq
This relation shows how the perturbative expansion of the $T$-matrix to any given order in the coupling constant cross-talks with lower-order terms and parts of those
will have increasingly higher inverse powers of $\hbar$. This is the origin of the eikonal
exponentiation in impact parameter space \cite{Cristofoli:2020uzm}. It is also the origin of the need to introduce the well-known Born subtractions, whether implemented by effective
field theory methods \cite{Cheung:2018wkq} or, equivalently, by solving the Lippmann-Schwinger equation associated with the corresponding relativistic
Hamiltonian \cite{Cristofoli:2019neg}.

Inspired by the different subtraction scheme behind the calculation of the conservative part to fourth Post-Minkowskian order of ref. \cite{Bern:2021dqo}, an alternative representation
of the $S$-matrix was suggested in ref. \cite{Damgaard:2021ipf}. In this representation, an Hermitian scattering matrix, denoted $N$, is introduced through
the operator identification
\beq
\hat{S} ~=~ \exp[i\hat{N}/\hbar] ~.
\label{Ndef}
\eeq
It was conjectured in ref. \cite{Damgaard:2021ipf} that two-to-two matrix elements of the operator $\hat{N}$, after a transform to impact-parameter space, yields
the radial action and hence, by simple differentiation, also the scattering angle. This was verified explicitly to third Post-Minkowskian order \cite{Damgaard:2021ipf}
and later checked, in the probe limit, up to fifth Post-Minkowskian order \cite{Bjerrum-Bohr:2021wwt}. More recently, the exponential representation has also
been checked against the fourth Post-Minkowskian order calculation of ref. \cite{Bern:2021dqo,Bern:2021yeh} for arbitrary masses \cite{Bjerrum-Bohr:2022ows}
but not including all radiation effects. There is thus substantial evidence that the exponential representation of the gravitational $S$-matrix captures the
classical dynamics of the conservative sector (and even parts of radiative effects) but a proof has so far still been lacking. One purpose of this paper is to provide such a proof.

Matrix elements of the exponential representation of the $S$-matrix resemble, after transforming to impact parameter space, the quantum field theoretic eikonal \cite{Collado:2018isu,KoemansCollado:2019ggb,DiVecchia:2019myk,
DiVecchia:2019kta,Bern:2020gjj,Parra-Martinez:2020dzs,DiVecchia:2022owy,DiVecchia:2022piu,Heissenberg:2022tsn}.
We stress, however, that these two representations are quite distinct beyond leading order. The $\hat{N}$-operator encapsulates by construction the semi-classical limit of the
$S$-matrix and its two-to-two matrix element is therefore expected to yield the corresponding radial action. Because $\hat{N}$ is already in the exponent there
are no superclassical contributions to it and all corrections to the radial action will be of quantum mechanical origin (and therefore not of interest here). The
$\hat{N}$-operator is thus more closely related to the WKB approximation than to the eikonal\footnote{For a recent comprehensive review of the eikonal formalism, see ref. \cite{DiVecchia:2023frv}.}.

Two other formalisms will be central to the understanding of gravitational two-body scattering in the Post-Minkowskian expansion. One is the KMOC formalism
\cite{Kosower:2018adc,Maybee:2019jus,Cristofoli:2021vyo,Cristofoli:2021jas,Herrmann:2021lqe,Herrmann:2021tct}, the other is the Post-Minkowskian worldline formalism 
\cite{Kalin:2020mvi,Kalin:2020fhe,Kalin:2020lmz,Mogull:2020sak,Jakobsen:2021smu,Liu:2021zxr,Dlapa:2021npj,Jakobsen:2021lvp,Jakobsen:2021zvh,Dlapa:2021vgp,Jakobsen:2022fcj,Jakobsen:2022psy,Kalin:2022hph,Dlapa:2022lmu,Dlapa:2023hsl,Jakobsen:2022zsx,Jakobsen:2023ndj}. The KMOC framework is, after appropriate reductions to the point-particle limit, intimately related
to the amplitude approach to gravitational scattering. Indeed some of the first resolutions of the puzzles at third Post-Minkowskian order came from expressing
KMOC observable in the form of cut amplitudes by reverse unitarity \cite{Herrmann:2021lqe,Herrmann:2021tct}. The worldline approach differs conceptually in
that the classical limit $\hbar \to 0$ can be taken from the outset, thus eliminating the need for subtractions altogether. In the end, the resulting integrals that must be
evaluated are nevertheless very similar and they are, not surprisingly, very closely related to the integrals that need to evaluated in the amplitude-based approach. It
becomes particularly clear in terms of the velocity cut method where the correspondence up to third Post-Minkowskian order has been shown to be one-to-one 
\cite{Bjerrum-Bohr:2021din}. This is not surprising in view of the fact that both formalisms amount to solving the classical Einstein field equations by Green function methods.

New issues have appeared at fourth Post-Minkowskian order of the gravitational expansion. These are related to both angular momentum loss and energy loss 
during the scattering process, losses which are due to the gravitationally radiated angular momentum and energy \cite{Manohar:2022dea}. 
There has been much progress on how to incorporate these effects in 
the eikonal formalism \cite{DiVecchia:2022owy,DiVecchia:2022piu,Heissenberg:2022tsn} but so far a complete computation has only been reported in work
using the worldline formalism \cite{Dlapa:2022lmu,Dlapa:2023hsl}. In order to tackle dissipation at this order, the worldline calculations have been rephrased in terms of 
the closed time paths of the Schwinger-Keldysh kind \cite{Jakobsen:2022psy,Kalin:2022hph}. This leads to a doubling of degrees of freedom, the use of
retarded (or advanced) propagators, and in general a much larger set of master integrals due to less symmetry of the integrands. 
It is interesting to contrast this with the KMOC formalism which
provides $S$-matrix expressions for the same quantities but based on standard amplitudes with Feynman propagators. In a recent paper \cite{Damgaard:2023vnx} we have
demonstrated the equivalence between the KMOC and worldline formulations in the classical limit. While this non-trivial relationship has been established on general grounds,
it is interesting that dissipative effects are accounted for quite differently in the two formulations due to the difference between Feynman and retarded/advanced propagators.

In this paper we combine the KMOC-formalism with the exponential representation of the $S$-matrix. We shall argue that such a combination is more economical than the 
conventional one based on the linear $T$-matrix representation of the $S$-matrix. It leads to very compact formulas for classical observables in gravity based on amplitudes and
it clarifies the inclusion of radiative effects in a simple diagrammatic fashion. Importantly, because the KMOC formalism makes no distinction between conservative and
dissipative contributions, classical observables are extracted in a universal manner from the matrix elements of the $\hat{N}$-operator by retaining all classical pieces.
As in the full amplitude computation at third Post-Minkowskian order~\cite{Bjerrum-Bohr:2021din} there is no need to separate different contributions. 
At any order in the expansion one only has to extract all classical terms of the matrix elements of $\hat{N}$ and derived quantities thereof. 

While equivalent to the worldline formulation in the Keldysh-Schwinger path integral, the formulas we shall present here have a structure that is straightforward to implement in 
terms of modern amplitude methods. Having different consistent formulations available is clearly an advantage and there is now a variety of approaches available for the
Post-Minkowskian expansion (see also refs.~\cite{Kalin:2019rwq,Bjerrum-Bohr:2019kec,Adamo:2022qci,Adamo:2022ooq,Brandhuber:2023hhy,Herderschee:2023fxh,Elkhidir:2023dco,Georgoudis:2023lgf,Gonzo:2023cnv}). This is particularly important
when the Post-Minkowskian expansion enters the new uncharted territory of higher orders.

We shall illustrate 
the simplicity of the combination of the $\hat{N}$-operator with the KMOC formalism by computing the full momentum kick (and hence scattering angle) to fourth 
Post-Minkowskian order. As we shall show, the required basis of master integrals is significantly smaller than that used in refs.~\cite{Dlapa:2022lmu,Dlapa:2023hsl} due
to the fact that we need only use Feynman propagators. Nevertheless, our results agree.

\section{The exponential representation of the gravitational $S$-matrix}\label{sec:Nformalism}

In this section we briefly review the exponential operator representation of the $S$-matrix. We first fix conventions. We consider
the Einstein-Hilbert action of two massive scalars (of masses $m_1$ and $m_2$) coupled to gravity,
\begin{equation}
S_{EH} = \int d^4 x \sqrt{-g} \Bigg[\frac{R}{16 \pi G}  + \frac12 \partial_\mu \phi_1\partial^\mu \phi_1 +\frac12 \partial_\mu \phi_2\partial^\mu \phi_2- {m_1^2\over2} \phi_1^2 - {m_2^2\over2} \phi_2^2\Bigg]\,.
\end{equation} 
Newton's constant is denoted by $G$ and $R$ is the Ricci 
scalar. We use a mostly-minus metric with flat Minkowski space at infinity, ${\rm diag}\,\eta_{\mu\nu}\equiv(1,-1,-1,-1)$, and expand the
full metric as $g_{\mu\nu}(x)\equiv \eta_{\mu\nu} + \sqrt{32 \pi G}h_{\mu\nu}(x)$.

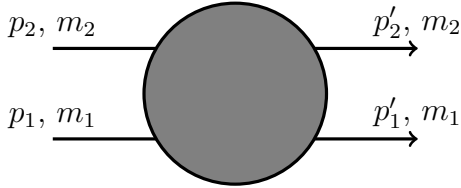
$$
\begin{tikzpicture}[scale=0.6]
\draw [black,very thick] [->] (-4,-1) to (4,-1);
\draw [black,very thick] [->] (-4,1) to (4,1);
\draw (-4,-1) node[above]{$p_1$, $m_1$};
\draw (4,-1) node[above]{$p'_1$, $m_1$};
\draw (-4,1) node[above]{$p_2$, $m_2$};
\draw (4,1) node[above]{$p'_2$, $m_2$};
\filldraw [color = black, fill=gray, very thick] (0,0) circle (2cm);
\end{tikzpicture}
$$

In this section we write everything in the standard language of {\em in-out} states and consider the two-to-two scattering with $p_1$ and
$p_2$ denoting incoming momenta and ${p_1'}$ and ${p_2'}$ outgoing momenta with
$p_1^2 = {p_1'}^2 = m_1^2$ and $p_2^2 =
{p_2'}^2 = m_2^2$.
In the center of mass frame with
\begin{equation}
  p_1=(E_1(p),\vec p), \qquad p_2=(E_2(p),-\vec p)
\end{equation}
we have
\begin{equation}\label{e:sdef} (p_1+p_2)^2 = ({p_1'}+{p_2'})^2 
= m_1^2+m_2^2+2m_1 m_2 \gamma,\quad \gamma \equiv \frac{p_1 \cdot p_2}{m_1 m_2}\,, \end{equation} 
\begin{equation}\label{e:tdef}
(p_1-{p_1'})^2 = ({p_2'}-p_2)^2\equiv q^2=-\vec{q}^{\,2}\,,\end{equation}

In ordinary scattering theory we wish to compute $S$-matrix elements. Here, instead, we shall focus on matrix elements of the Hermitian
operator $\hat{N}$ defined by eq. (\ref{Ndef}), in particular, for two-to-two scattering, 
\begin{equation}\label{e:N4pt}
N(\gamma,q^2)=\langle p_1',p_2' \vert \hat{N} \vert p_1,p_2\rangle ~.
\end{equation}
This should be contrasted with the standard Born expansion of the $S$-matrix based on
\begin{equation}
  \hat{S}=1 + \frac{i}{\hbar}\hat{T}  
  \label{e:Born}
\end{equation}
and the usual scattering amplitude $M(p_1,p_2,p_1',p_2') $ defined by
\begin{equation}
  \langle p_1',p_2'| \hat{T} | p_1, p_2 \rangle  ~=~  (2\pi \hbar)^D\delta^{(D)}(p_1+p_2-p_1'-p_2') M(p_1,p_2,p_1',p_2')  ~,
\end{equation}
in dimensions $D=4-2\epsilon$.
As detailed in ref.~\cite{Damgaard:2021ipf} it is straightforward to expand the exponential representation and derive the infinite
sequence of relations between operators $\hat{N}$ and
$\hat{T}$ in perturbation theory. In the two-to-two sector the operators have perturbative expansions that we can write compactly as
\begin{eqnarray}\label{e:TNexp}
\hat{T}  &=&  G\hat{T}_0 + G^{3/2}\hat{T}_0^{\rm rad} + G^2\hat{T}_1 + G^{5/2}\hat{T}_1^{\rm rad} + G^3\hat{T}_2 + \cdots \cr
\hat{N}  &=&  G\hat{N}_0 + G^{3/2}\hat{N}_0^{\rm rad} + G^2\hat{N}_1 + G^{5/2}\hat{N}_1^{\rm rad} + G^3\hat{N}_2 + \cdots
\end{eqnarray}
from which we straightforwardly can solve for $G$'s in terms of $T$'s by expanding the exponential.
Integer powers of $G$ describe interactions with an even number of
graviton vertices while half-integer powers describe interactions with an odd number of gravitons. The separation of 
operators with superscript rad refers only to the associated half-integer power of $G$. We find it useful diagrammatically
to make this distinction (see also below) but it has no further meaning beyond this. There are clearly also radiative terms in the
even powers.

At order $G^4$ the relation reads
\begin{multline}\label{e:NT3}
             {\hat{N}}_3  =  {\hat{T}}_3 -
                            \frac{i}{2\hbar}(\hat{N}^{\rm
                            rad}_1\hat{N}^{\rm rad}_0+\hat{N}^{\rm
                            rad}_0\hat{N}^{\rm rad}_1)-
                            \frac{i}{2\hbar}{\hat{T}}_1^2 -
                            \frac{i}{2\hbar}(\hat{T}_0
                            \hat{T}_2+\hat{T}_2 \hat{T}_0)\cr
   -  \frac{1}{12\hbar^2} [\hat{N}_0^{\text{rad}},[\hat{N}_0^{\text{rad}},\hat{N}_0]] -
                            \frac{1}{3\hbar^2}(\hat{T}_0^2
                            \hat{T}_1+\hat{T}_0 \hat{T}_1\hat{T}_0 +
                            \hat{T}_1\hat{T}_0^2) + \frac{i}{4
                            \hbar^3}\hat{T}_0^4\,.
                        \end{multline} 
and it is elementary to generalize this to higher orders. Note that we have combined some of the $T$-matrices into $N$-matrices on the right hand
side, thus making the cancellation among superclassical pieces associated with those manifest. This also aids in understanding the separation into real and
imaginary parts. We remind that $\hat{N}$ is Hermitian so that two-to-two scalar matrix elements of that operator are real.

The obvious way to evaluate matrix elements of the $\hat N$ operator by conventional field theory methods is
to insert a complete set of momentum eigenstates between all products of $T$-matrices and truncate to the desired order in $G$. 
Then matrix elements can be evaluated by
standard Feynman rules of scattering theory. Here the complete set of states is spanned by
two massive scalar particles: one of momentum $k_1$ and mass
$m_1$, the other of momentum $k_2$ and mass $m_2$, together with  any number $n$ of
massless gravitons. We denote such states by $ |k_1,k_2;\ell_1,\dots,\ell_n\rangle$.
These states are normalized relativistically according to
\begin{multline}
  \langle
  k_1,k_2;\ell_1,\dots,\ell_n|k'_1,k'_2;\ell'_1,\dots,\ell'_m\rangle =
  \delta_{n,m} \prod_{i=1}^2 2E_{k_i}
  (2\pi\hbar)^{D-1}\delta^{(D-1)}(k_i-k'_i)\cr
  \times\prod_{i=1}^n 2E_{\ell_i} \,(2\pi\hbar)^{D-1} \delta^{(D-1)}(\ell_i-\ell_i')  ,
\end{multline}
and the completeness relation is given by
\begin{equation}\label{e:complete}
1  =  \sum_{n=0}^{\infty}\frac{1}{n!}\int \prod_{i=1}^2
d\mathrm{\Pi}_{k_{i}} \prod_{r=1}^n d\mathrm{\Pi}_{\ell_{r}}
|k_1,k_2;\ell_1,\dots,\ell_n\rangle   \langle k_1, k_2; \ell_1, \ldots \ell_n |.
\end{equation}
including a sum over graviton helicities.
Here $d\mathrm{\Pi}$ is the standard Lorentz invariant phase space measure, $i.e.$,
\begin{equation}\label{e:Pimassive}
d\mathrm{\Pi}_{k_{i}}=
\frac{d^Dk_i}{(2\pi\hbar)^{D-1}}\delta^+((k_i)^2-m_i^2) = \frac{d^Dk_i}{(2\pi\hbar)^{D-1}}\theta(k_i^0)\delta((k_i)^2-m_i^2)
\qquad \textrm{for}\qquad i=1,2
\end{equation}
for the massive states, and similarly for the massless gravitons.

We now insert the completeness relation between all operator
products to get the three-loop relation between matrix elements of the
$\hat N$ and the $\hat T$ operators
\begin{equation}
       \langle p_1',p_2'|       {\hat{N}}_3 |p_1,p_2\rangle = \langle p_1',p_2'|   {\hat{T}}_3 |p_1,p_2\rangle+L_0+L_1+L_2
     \end{equation}
 which we then expand in powers of $G$. Keeping track of this overall power of $G$, we can view it as an expansion
 in the number of gravitons connecting the operators. First, with just the massive states inserted,
\begin{multline}
L_0= -\frac{i}{2}\begin{gathered}
\begin{tikzpicture}[scale=0.5]
\draw [black,very thick] [->] (-4,-.8) to (4,-.8);
\draw [black,very thick] [->] (-4,.8) to (4,.8);
\draw [red] (0,-1) to (0,-.6);
\draw [red] (0,1) to (0,.6);
\filldraw [color = black, fill=white, very thick] (-2,0) circle (1cm)
node {$T_0$};
\filldraw [color = black, fill=white, very thick] (2,0) circle (1cm)
node {$T_2$};
\end{tikzpicture}
\end{gathered}
 -\frac{i}{2}
 \begin{gathered}
\begin{tikzpicture}[scale=0.5]
\draw [black,very thick] [->] (-4,-.8) to (4,-.8);
\draw [black,very thick] [->] (-4,.8) to (4,.8);
\draw [red] (0,-1) to (0,-.6);
\draw [red] (0,1) to (0,.6);
\filldraw [color = black, fill=white, very thick] (-2,0) circle (1cm)
node {$T_2$};
\filldraw [color = black, fill=white, very thick] (2,0) circle (1cm)
node {$T_0$};
\end{tikzpicture}
\end{gathered}
-\frac{i}{2}
 \begin{gathered}
\begin{tikzpicture}[scale=0.5]
\draw [black,very thick] [->] (-4,-.8) to (4,-.8);
\draw [black,very thick] [->] (-4,.8) to (4,.8);
\draw [red] (0,-1) to (0,-.6);
\draw [red] (0,1) to (0,.6);
\filldraw [color = black, fill=white, very thick] (-2,0) circle (1cm)
node {$T_1$};
\filldraw [color = black, fill=white, very thick] (2,0) circle (1cm)
node {$T_1$};
\end{tikzpicture}
\end{gathered}\cr
+\frac{i}{4}\begin{gathered}
\begin{tikzpicture}[scale=0.5]
\draw [black,very thick] [->] (-8,-.8) to (8,-.8);
\draw [black,very thick] [->] (-8,.8) to (8,.8);
\draw [red] (-4,-1) to (-4,-.6);
\draw [red] (-4,1) to (-4,.6);
\draw [red] (0,1) to (0,.6);
\draw [red] (0,-1) to (0,-.6);
\draw [red] (4,-1) to (4,-.6);
\draw [red] (4,1) to (4,.6);
\filldraw [color = black, fill=white, very thick] (-6,0) circle (1cm)
node {$T_0$};
\filldraw [color = black, fill=white, very thick] (-2,0) circle (1cm)
node {$T_0$};
\filldraw [color = black, fill=white, very thick] (2,0) circle (1cm)
node {$T_0$};
\filldraw [color = black, fill=white, very thick] (6,0) circle (1cm)
node {$T_0$};
\end{tikzpicture}
\end{gathered}-\frac{1}{3} \begin{gathered}
\begin{tikzpicture}[scale=0.5]
\draw [black,very thick] [->] (-8,-.8) to (4,-.8);
\draw [black,very thick] [->] (-8,.8) to (4,.8);
\draw [red] (-4,-1) to (-4,-.6);
\draw [red] (-4,1) to (-4,.6);
\draw [red] (0,1) to (0,.6);
\draw [red] (0,-1) to (0,-.6);
\filldraw [color = black, fill=white, very thick] (-6,0) circle (1cm)
node {$T_0$};
\filldraw [color = black, fill=white, very thick] (-2,0) circle (1cm)
node {$T_0$};
\filldraw [color = black, fill=white, very thick] (2,0) circle (1cm)
node {$T_1$};
\end{tikzpicture}
\end{gathered}\\ -\frac{1}{3} \begin{gathered}
\begin{tikzpicture}[scale=0.5]
\draw [black,very thick] [->] (-8,-.8) to (4,-.8);
\draw [black,very thick] [->] (-8,.8) to (4,.8);
\draw [red] (-4,-1) to (-4,-.6);
\draw [red] (-4,1) to (-4,.6);
\draw [red] (0,1) to (0,.6);
\draw [red] (0,-1) to (0,-.6);
\filldraw [color = black, fill=white, very thick] (-6,0) circle (1cm)
node {$T_0$};
\filldraw [color = black, fill=white, very thick] (-2,0) circle (1cm)
node {$T_1$};
\filldraw [color = black, fill=white, very thick] (2,0) circle (1cm)
node {$T_0$};
\end{tikzpicture}
\end{gathered}-\frac{1}{3} \begin{gathered}
\begin{tikzpicture}[scale=0.5]
\draw [black,very thick] [->] (-8,-.8) to (4,-.8);
\draw [black,very thick] [->] (-8,.8) to (4,.8);
\draw [red] (-4,-1) to (-4,-.6);
\draw [red] (-4,1) to (-4,.6);
\draw [red] (0,1) to (0,.6);
\draw [red] (0,-1) to (0,-.6);
\filldraw [color = black, fill=white, very thick] (-6,0) circle (1cm)
node {$T_1$};
\filldraw [color = black, fill=white, very thick] (-2,0) circle (1cm)
node {$T_0$};
\filldraw [color = black, fill=white, very thick] (2,0) circle (1cm)
node {$T_0$};
\end{tikzpicture}
\end{gathered}
\end{multline}
Next, with the inclusion of one graviton,

\begin{multline}
 L_1= -\frac{i}{2}\begin{gathered}
\begin{tikzpicture}[scale=0.5]
\draw [black,very thick] [->] (-8,-.8) to (0,-.8);
\draw [black,very thick] [->] (-8,.8) to (0,.8);
\draw [black,snake it]  (-5,0) to (-2,0);
\draw [red] (-4,-1) to (-4,-.6);
\draw [red] (-4,1) to (-4,.6);
\draw [red] (-4,.3) to (-4,-.3);
\filldraw [color = black, fill=white, very thick] (-6,0) circle (1cm)
node {$N_0^{\rm rad}$};
\filldraw [color = black, fill=white, very thick] (-2,0) circle (1cm)
node {$N_1^{\rm rad}$};
\end{tikzpicture}
\end{gathered} -\frac{i}{2} \begin{gathered}
\begin{tikzpicture}[scale=0.5]
\draw [black,very thick] [->] (-8,-.8) to (0,-.8);
\draw [black,very thick] [->] (-8,.8) to (0,.8);
\draw [black,snake it]  (-5,0) to (-2,0);
\draw [red] (-4,-1) to (-4,-.6);
\draw [red] (-4,1) to (-4,.6);
\draw [red] (-4,.3) to (-4,-.3);
\filldraw [color = black, fill=white, very thick] (-6,0) circle (1cm)
node {$N_1^{\rm rad}$};
\filldraw [color = black, fill=white, very thick] (-2,0) circle (1cm)
node {$N_0^{\rm rad}$};
\end{tikzpicture}
\end{gathered}\cr -\frac{1}{12}\begin{gathered}
\begin{tikzpicture}[scale=0.5]
\draw [black,very thick] [->] (-8,-.8) to (4,-.8);
\draw [black,very thick] [->] (-8,.8) to (4,.8);
\draw [black,snake it]  (-5,0) to (-2,0);
\draw [red] (-4,-1) to (-4,-.6);
\draw [red] (-4,1) to (-4,.6);
\draw [red] (0,1) to (0,.6);
\draw [red] (0,-1) to (0,-.6);
\draw [red] (-4,.3) to (-4,-.3);
\filldraw [color = black, fill=white, very thick] (-6,0) circle (1cm)
node {$N_0^{\rm rad}$};
\filldraw [color = black, fill=white, very thick] (-2,0) circle (1cm)
node {$N_0^{\rm rad}$};
\filldraw [color = black, fill=white, very thick] (2,0) circle (1cm)
node {$N_0$};
\end{tikzpicture}
\end{gathered}+ \frac{1}{6}
\begin{gathered}\vspace{.3cm}
\begin{tikzpicture}[scale=0.5]
\draw [black,very thick] [->] (-8,-.8) to (4,-.8);
\draw [black,very thick] [->] (-8,.8) to (4,.8);
\draw [black,snake it]  (-2,1.5) to (2,1);
\draw [black,snake it]  (-2,1.5) to (-6,1);
\draw [red] (-4,-1) to (-4,-.6);
\draw [red] (-4,1) to (-4,.6);
\draw [red] (0,1) to (0,.6);
\draw [red] (0,-1) to (0,-.6);
\draw [red] (-2,1.8) to (-2,1.2);
\filldraw [color = black, fill=white, very thick] (-6,0) circle (1cm)
node {$N_0^{\rm rad}$};
\filldraw [color = black, fill=white, very thick] (-2,0) circle (1cm)
node {$N_0$};
\filldraw [color = black, fill=white, very thick] (2,0) circle (1cm)
node {$N_0^{\rm rad}$};
\end{tikzpicture}
\end{gathered}\cr-\frac{1}{12}
\begin{gathered}
\begin{tikzpicture}[scale=0.5]
\draw [black,very thick] [->] (-8,-.8) to (4,-.8);
\draw [black,very thick] [->] (-8,.8) to (4,.8);
\draw [black,snake it]  (-1.5,0) to (1.5,0);
\draw [red] (-4,-1) to (-4,-.6);
\draw [red] (-4,1) to (-4,.6);
\draw [red] (0,1) to (0,.6);
\draw [red] (0,-1) to (0,-.6);
\draw [red] (0,.3) to (0,-.3);
\filldraw [color = black, fill=white, very thick] (-6,0) circle (1cm)
node {$N_0$};
\filldraw [color = black, fill=white, very thick] (-2,0) circle (1cm)
node {$N_0^{\rm rad}$};
\filldraw [color = black, fill=white, very thick] (2,0) circle (1cm)
node {$N_0^{\rm rad}$};
\end{tikzpicture}
\end{gathered}
\end{multline}
as well as one graviton inserted twice:
\begin{equation}
  \label{e:L3}
  L_2=\frac{1}{6}\begin{gathered}\vspace{.3cm}
\begin{tikzpicture}[scale=0.5]
\draw [black,very thick] [->] (-8,-.8) to (4,-.8);
\draw [black,very thick] (-8,.8) to (-6,.8);
\draw [black,very thick] [->] (2,.8) to (4,.8);
\draw [black,snake it] [->] (-6,.8) to (2,.8);
\draw [black,very thick]  (-2,1.5) to (2,1);
\draw [black, very thick]  (-2,1.5) to (-6,1);
\draw [red] (-4,-1) to (-4,-.6);
\draw [red] (-4,1) to (-4,.6);
\draw [red] (0,1) to (0,.6);
\draw [red] (0,-1) to (0,-.6);
\draw [red] (-2,1.8) to (-2,1.2);
\filldraw [color = black, fill=white, very thick] (-6,0) circle (1cm)
node {$N_0^{\rm rad}$};
\filldraw [color = black, fill=white, very thick] (-2,0) circle (1cm)
node {$N_0$};
\filldraw [color = black, fill=white, very thick] (2,0) circle (1cm)
node {$N_0^{\rm rad}$};
\end{tikzpicture}
\end{gathered}+\frac{1}{6}\begin{gathered}\vspace{-.5cm}
\begin{tikzpicture}[scale=0.5]
\draw [black,very thick] [->] (-8,.8) to (4,.8);
\draw [black,very thick] (-8,-.8) to (-6,-.8);
\draw [black,very thick] [->] (2,-.8) to (4,-.8);
\draw [black,snake it] [->] (-6,-.8) to (2,-.8);
\draw [black,very thick]  (-2,-1.5) to (2,-1);
\draw [black, very thick]  (-2,-1.5) to (-6,-1);
\draw [red] (-4,-1) to (-4,-.6);
\draw [red] (-4,1) to (-4,.6);
\draw [red] (0,1) to (0,.6);
\draw [red] (0,-1) to (0,-.6);
\draw [red] (-2,-1.8) to (-2,-1.2);
\filldraw [color = black, fill=white, very thick] (-6,0) circle (1cm)
node {$N_0^{\rm rad}$};
\filldraw [color = black, fill=white, very thick] (-2,0) circle (1cm)
node {$N_0$};
\filldraw [color = black, fill=white, very thick] (2,0) circle (1cm)
node {$N_0^{\rm rad}$};
\end{tikzpicture}
\end{gathered}
\end{equation}
Note that the completeness relation enforces the inclusion of graph topologies that are partly
disconnected, such as the graviton line skipping one internal operator as well as the Compton-type
contributions in the last line where scalars skip an internal operator. Such intermediate states begin to contribute for the first time at fourth
Post-Minkowskian order because up to and including third Post-Minkowskian order they have no support on physical
kinematics. To fourth order in $G$ no further insertions of graviton states are possible when evaluating $N$-matrix elements through use of
eq. (\ref{e:NT3}).

Although written as an apparent expansion in $1/\hbar$ one must keep in mind
that additional factors of $\hbar$ (of both positive and negative
powers) arise when computing matrix elements. Since matrix elements of the $\hat{N}$ are manifestly free of superclassical contributions, the subtractions on the right
hand side of eq.~\eqref{e:NT3} ensure cancellations among
all superclassical terms arising from the $\hat T$-matrix, here including order $1/\hbar^3$-terms.
We shall show in section~\ref{sec:KMOCclassical}  how this implies the
cancellation of the superclassical terms when evaluating observables
in the KMOC formalism.

One advantage of the exponential representation is that we can ignore these superclassical cancellations that are guaranteed to occur anyway and thus focus
exclusively on the pieces that have a well-defined $\hbar \to 0$ limit. The systematic way to extract this classical limit of matrix elements of the $\hat{N}$-operator is
by means of velocity cuts. This will be described next.

\subsection{The classical limit and velocity cuts}

The notion of velocity
cuts~\cite{Bjerrum-Bohr:2021vuf,Bjerrum-Bohr:2021din,Bjerrum-Bohr:2021wwt}
is computationally useful for extracting the
classical limit. The basic idea is to combine massive propagator lines in pairs, each having denominators
that are linear in the external momenta but with opposite signs, thus effectively reducing to delta-function
constraints that are linear in momenta. Ignoring soft momentum corrections, this puts the massive lines on-shell and removes one momentum integration, 
thus enforcing the first link to the classical worldline
formalism. 

The classical limit $\hbar\to0$ of the massive amplitude is
obtained by scaling the momentum transfer $q=\hbar \underline
q$ with $\underline q$ fixed, and scaling the loop integration momenta
$\ell_i=\hbar|\underline q|\, \bar\ell_i$. The amplitude will involve two
massive propagators,  
\begin{equation}
  {1\over \left(\ell+p_r\right)^2-m_r^2+i\varepsilon}  ={1\over
    2\ell\cdot p_r+\ell^2+i\varepsilon} \qquad r=1,2
\end{equation}
where $\ell$ is a generic loop momentum.
In the classical limit we have
\begin{equation}
  {1\over 2 \hbar |\underline q| \ell\cdot p_r+ \hbar^2|\underline
    q|^2 \ell^2+i\varepsilon}\simeq {1\over2 \hbar |\underline q| } {1\over \ell\cdot p_r+ i\varepsilon},
\end{equation}
so that the $\ell^2$ part is subleading and the
massive propagators effectively become linear.  Combinations of such
linear propagators using
\begin{equation}
 \lim_{\varepsilon\to0} \left(  {1\over
    2\ell\cdot p_r+\ell^2+i\varepsilon} + {1\over
    2\ell\cdot p_r-\ell^2+i\varepsilon} \right) =-2i\pi \delta(2\ell\cdot p_r)
\end{equation}
lead to delta-function insertions in the loops.  The higher
order $O(\hbar^2\underline q^2)$ pieces do not contribute to
the classical result, thus eventually making the link to the classical worldline formalism, as we shall discuss below.
The classical part of the massive
two-to-two amplitude at $L$-loop order has exactly $L$ velocity
cuts~\cite{Bjerrum-Bohr:2021wwt}. Therefore the classical amplitude
can be reduced on a special class of
Post-Minkowskian master integrals with $L$ such delta-function insertions.

This set of master integrals also arises from the worldline
formalism~\cite{Kalin:2020fhe,Dlapa:2021vgp} 
as explained at two-loop order  in ref.~\cite{Bjerrum-Bohr:2021din}. 
An alternative approach is based on the heavy-mass expansion of scattering amplitudes \cite{Damgaard:2019lfh,Aoude:2020onz}. 
The classical terms 
can be re-organized in terms of a heavy mass expansion rather than as the $\hbar \to 0$ viewpoint taken here. The result is an effective field theory of 
linearized massive propagators and, in loops, precisely corresponding to the velocity cuts 
\cite{Brandhuber:2021eyq,Brandhuber:2021bsf,Brandhuber:2022enp}.

\section{The KMOC formalism and the exponential representation}\label{sec:KMOCclassical}

The KMOC formalism as originally defined in~\cite{Kosower:2018adc} considers an initial {\em in}-state of two massive scalars at time $t=-\infty$,
\begin{equation}
    \left.|\text{in}\right\rangle =
\int{d{\mathrm{\Pi }}_{p_{1}}d{\mathrm{\Pi }}_{p_{2}}}{\mathrm{\tilde\Phi_1}}(p_1)
{\mathrm{\tilde\Phi_2 }}(p_2)e^{\frac{i}{\hbar}bp_1}
\left.|p_1,p_2;0\right\rangle
\end{equation}
where the state $\left.|p_1,p_2;0\right\rangle$ is a momentum eigenstate of two massive scalars and the ``0'' indicates that there is no radiation
present at $t = -\infty$.
In the classical limit the wavefunctions $\tilde{\mathrm{\Phi }}(p_i)$ are chosen so as to represent two
localized scalars separated by impact parameter $b^{\mu}$. 
A complete set of states containing an arbitrary number of gravitons is as described in 
eq. (\ref{e:complete}) but the initial state at $t=-\infty$ is taken to be free of gravitons, as shown.

A change in an observable corresponding to an operator $\hat{O}$ from $t = -\infty$ to $t = +\infty$ is then \cite{Kosower:2018adc},
\begin{equation}
\langle \Delta \hat{O} \rangle = \left\langle\text{in}\right\vert\hat{S}^{\dagger} \hat{O}\hat{S}\left\vert\text{in}\right\rangle
- \left\langle\text{in}\right\vert\hat{O}\left\vert\text{in}\right\rangle = 
\left\langle\text{in}\right\vert\hat{S}^{\dagger} [\hat{O},\hat{S}]\left\vert\text{in}\right\rangle ~.
\label{DeltaOexpec}
\end{equation}
Using the linear Born representation of the $S$-matrix~\eqref{e:Born}
leads to the KMOC formula 
\begin{equation}
\langle \Delta \hat{O} \rangle ={i\over\hbar}
\left\langle\text{in}\right\vert[\hat{O},\hat{T}]\left\vert\text{in}\right\rangle
+{1\over \hbar^2}
 \left\langle\text{in}\right\vert\hat{T}^{\dagger}[\hat{O},\hat{T}]\left\vert\text{in}\right\rangle
\label{KMOCT}
\end{equation}
In the small $\hbar$ limit this expression leads to the evaluation of the change in a
classical observable after the delicate cancellations of superclassical terms.

Here we instead explore consequences of using the exponential
representation of the $S$-matrix. This will lead to a simple and
efficient way to extract the change in a classical observable, including dissipative effects.

In an alternative viewpoint we consider the change $\Delta \hat{O}$ of
an operator $\hat{O}$ from $t=-\infty$ to $t=+\infty$ as
\begin{equation}
\Delta \hat{O}=\hat{S}^{\dagger} \hat{O}\hat{S}- \hat{O} ~.
\label{DeltaOdef}
\end{equation}
which then has to be evaluated between {\em in}-states of $t=-\infty$.
Inserting the exponential representation of the $\hat S$ operator of
eq.~\eqref{Ndef} together with the crucial property of Hermiticity of $\hat{N}$, 
\begin{equation}
\Delta \hat{O}=e^{-{i\hat{N}\over\hbar}} \hat{O}e^{i\hat{N}\over \hbar} - \hat{O} ~.
\label{DeltaOdefexp}
\end{equation}
allows us to rewrite 
eq.~\eqref{DeltaOdefexp} by means of the Campbell identity that expands the two exponentials as an infinite
sum of nested commutators, 
\begin{equation}\label{e:DeltaOcomm}
\Delta \hat{O}=\sum_{n \geq 1} \frac{(-i)^n}{\hbar^{n}n!} \underbrace{[\hat{N},[\hat{N},\dots,[\hat{N},\hat{O}]]]}_{\text{n times}}.
\end{equation}
This rewriting, which is where we use unitarity of the $S$-matrix, will play a crucial role in our all-order proofs
because it displays the iterative structure of the KMOC formalism when combined with the exponential representation.
It is convenient to define 
\begin{equation}
    \hat{A}_n^{\hat{O}} \equiv \frac{1}{\hbar^n}\underbrace{[\hat{N},[\hat{N},\dots,[\hat{N},\hat{O}]]]}_{\text{n times}} .
\end{equation}
The nested commutator structure implies the operator relation
\begin{equation}\label{e:AOn}
\hat{A}_n^{\hat{O}}=\hat{A}_{1}^{\hat{A}_{n-1}^{\hat{O}}}=\hat{A}_{1}^{\hat{A}_{1}^{\rddots
    {\hat{A}_1^{\hat{O}}}}}.
\end{equation}
Importantly, when we evaluate matrix elements by means of insertions of complete sets of states, this iterative structure is preserved (since all
we do is to insert factors of unity).

Repeating the steps described in ref. \cite{Kosower:2018adc}, we can insert the above expression in the KMOC-expression
and take the limit
of localized massive states. The result is
\begin{equation}
\langle\Delta \hat{O}\rangle (p_1,p_2,b)= \int
\frac{d^Dq}{(2\pi)^{D-2}}\delta(2p_1\cdot q - q^2)\delta(2p_2\cdot q +
q^2)e^{i{b\cdot q\over \hbar}}
\langle p_1'p_2' | \Delta O | p_1 p_2\rangle
\label{KMOCbspace}
\end{equation}
where $p_1' = p_1 - q$ and $p_2' = p_2 + q$. In this form it is clear that a first step is the evaluation of the
matrix element $\langle p_1'p_2' | \Delta O | p_1 p_2\rangle$, followed by the shown Fourier transform to $b$-space.

One noticeable feature of the KMOC-formalism for (non-spinning) black-hole scattering is that it always entails the evaluation
of matrix elements of an operator (\ref{DeltaOdef}) between two-particle scalar states.
For an observable corresponding to an Hermitian operator $\hat O$ the corresponding 
$\Delta O$ is clearly Hermitian as well. Two-particle scalar matrix elements of this $\Delta O$ are then real, as follows from 
time-reversal symmetry. The reality of the expectation value is preserved by the insertion of the completeness relation since it
just amounts to the insertion of factors of unity.

\subsection{Cancellation of superclassical terms: the conservative sector}\label{sec:conservative}

In this section we first show how the $N$-operator formalism provides a
simple way to demonstrate the cancellation of the superclassical pieces when restricted to the conservative sector.  We next
give a general formula valid to all orders in $G$
for a general operator in section~\ref{sec:scalarcons} and a
vector operator in section~\ref{sec:vectorcons}. The application to
the momentum kick $\Delta P_1$ is pursued in section~\ref{sec:DeltaP1cons}.

\subsubsection{The classical limit}\label{sec:scalarcons}
We start with a general operator $\hat O$ and consider the term with $n=1$ in~\eqref{e:DeltaOcomm}
\begin{equation}\label{e:AOdef}
  \mathcal A_1^O(p_1,p_2,q)={1\over \hbar} \langle p_1',p_2'| [\hat N, \hat O]|p_1,p_2\rangle  
\end{equation}
and 
we first analyze the conservative case where gravitons are not included in the set of inserted on-shell states.
This is graphically represented as
$$
\mathcal A_1^O(p_1,p_2,q)|^{\rm cons.}= 
 \begin{gathered}
\begin{tikzpicture}[scale=0.5]
\draw [black,very thick] [->] (-4,-.8) to (4,-.8);
\draw [black,very thick] [->] (-4,.8) to (4,.8);
\draw [red] (0,-1) to (0,-.6);
\draw [red] (0,1) to (0,.6);
\filldraw [color = black, fill=white, very thick] (-2,0) circle (1cm)
node {$N$};
\filldraw [color = black, fill=white, very thick] (2,0) circle (1cm)
node {$O$};
\end{tikzpicture}
\end{gathered}
-
 \begin{gathered}
\begin{tikzpicture}[scale=0.5]
\draw [black,very thick] [->] (-4,-.8) to (4,-.8);
\draw [black,very thick] [->] (-4,.8) to (4,.8);
\draw [red] (0,-1) to (0,-.6);
\draw [red] (0,1) to (0,.6);
\filldraw [color = black, fill=white, very thick] (-2,0) circle (1cm)
node {$O$};
\filldraw [color = black, fill=white, very thick] (2,0) circle (1cm)
node {$N$};
\end{tikzpicture}
\end{gathered}
$$
where the red line indicates where we insert the intermediate two-particle state,
corresponding to 
\begin{multline}
  \mathcal A_1^O(p_1,p_2,q)= \frac{1}{\hbar}\int d\Pi_{q_1} d\Pi_{q_2}\Big(\langle
  p_1',p_2'| \hat N|q_1,q_2\rangle\langle q_1,q_2 |\hat O|p_1,p_2\rangle \cr- \langle
  p_1',p_2'| \hat O|q_1,q_2\rangle\langle q_1,q_2 |\hat N|p_1,p_2\rangle \Big).
\end{multline}
It is convenient to factor out overall energy-momentum conservation and write
\begin{equation}
    \langle
  p_1',p_2'| \hat N|p_1,p_2\rangle= N(\gamma ,q^2) (2\pi\hbar)^D \delta(p_1'+p'_2-p_1-p_2)
\end{equation}
and
\begin{equation}
    \langle
  p_1',p_2'| \hat O|p_1,p_2\rangle= O(p_1,p_2,q) (2\pi\hbar)^D \delta(p'_1+p'_2-p_1-p_2).
\end{equation}
We can use one of the energy-momentum conservation delta-functions to remove integration variable $q_2$
After defining $k_1=q_1-p_1$ and
using the scaled momenta $\underline q$ and
$\underline k_1$ such that
$p_1'=p_1-q=p_1-\hbar \underline{q} $, $p_2'=p_2+q=p_2+\hbar
\underline{q}$ we change variables to get
\begin{multline}
\mathcal A_1^O(p_1,p_2,q)=\hbar\int \frac{ d^D \underline{k}_1}{(2\pi)^{D-2}} \delta^+((p_1+\hbar
\underline{k}_1)^2-m_1^2)\delta^+((p_2-\hbar \underline{k}_1)^2-m_2^2)\cr
\times\Big(N(\gamma,\hbar^2 (\underline k_1+\underline q)^2)
O(p_1,p_2,-\hbar \underline{k}_1) - O(p_1+\hbar \underline{k}_1,p_2-\hbar
\underline{k}_1,\hbar(\underline{q}+\underline{k}_1))N(\gamma,\hbar^2
\underline{k}_1^2) \Big)\cr
\times (2\pi \hbar)^D \delta(p_1+p_2-p_1'-p_2').
\end{multline}
Setting
\begin{equation}
  \mathcal A_1^O(p_1,p_2,q)= A_1^O(p_1,p_2,q)  (2\pi \hbar)^D \delta(p_1+p_2-p_1'-p_2').
\end{equation}
Changing variables $\underline{k}_1 \rightarrow -\underline{k}_1-\underline{q}$ to the
second term of the sum gives
\begin{multline}
A_1^O(p_1,p_2,q)=\hbar\int \frac{ d^D \underline{k}_1}{(2\pi)^{D-2}}O(p_1,p_2,-\hbar \underline{k}_1) N(\gamma,\hbar^2 (\underline{k}_1+\underline{q})^2) \cr\times\delta^+((p_1+\hbar \underline{k}_1)^2-m_1^2)\delta^+((p_2-\hbar \underline{k}_1)^2-m_2^2)\cr - \hbar\int \frac{ d^D \underline{k}_1}{(2\pi)^{D-2}} O(p_1-\hbar (\underline{k}_1+\underline{q}),p_2+\hbar (\underline{k}_1+\underline{q}),-\hbar \underline{k}_1)N(\gamma,\hbar^2 (\underline{k}_1+\underline{q})^2) \cr\times \delta^+((p_1-\hbar (\underline{k}_1+\underline{q}))^2-m_1^2)\delta^+((p_2+\hbar(\underline{k}_1+\underline{q}))^2-m_2^2).
\end{multline}
Doing the small $\hbar$ expansion of the integrand leads to
\begin{multline}
O(p_1,p_2,-\hbar \underline{k}_1)\delta^+((p_1+\hbar
\underline{k}_1)^2-m_1^2)\delta^+((p_2-\hbar \underline{k}_1)^2-m_2^2)\cr
-O(p_1-\hbar (\underline{k}_1+\underline{q}),p_2+\hbar (\underline{k}_1+\underline{q}),-\hbar \underline{k}_1)\delta^+((p_1-\hbar (\underline{k}_1+\underline{q}))^2-m_1^2)\delta^+((p_2+\hbar(\underline{k}_1+\underline{q}))^2-m_2^2)\cr
=\frac{2}{\hbar} ((\underline{k}_1+\underline{q}) \cdot \underline{k}_1)O(p_1,p_2,-\hbar \underline{k}_1)\Big((\delta^+)'(2 p_1 \cdot \underline{k}_1)\delta^+(-2 p_2 \cdot \underline{k}_1)+\delta^+(2 p_1 \cdot \underline{k}_1)(\delta^+)'(-2 p_2 \cdot \underline{k}_1)\Big)\\+\frac{1}{\hbar} (\underline{k}_1^\mu+\underline{q}^\mu)(\nabla^\mu O(p_1,p_2,-\hbar \underline{k}_1))\delta^+(2 p_1 \cdot \underline{k}_1)\delta^+(-2 p_2 \cdot \underline{k}_1).
\end{multline}
where we have introduced the derivative 
\begin{equation}\label{e:NablaF}
  \nabla_\mu [\mathcal F]\equiv\frac{\partial \mathcal
    F}{\partial p_1^\mu}-\frac{\partial \mathcal F}{\partial p_2^\mu}.
  \end{equation}
Consequently the $\hbar$ expansion of $A_1^O$ takes the form 
\begin{multline}
A_1^O(p_1,p_2,q)=\int \frac{d^D \underline{k}_1}{(2\pi)^{D-2}} N(\gamma,\hbar^2 (\underline{k}_1+\underline{q})^2)\cr\times (\underline{k}_1^\mu+\underline{q}^\mu)\nabla_\mu \Big(O(p_1,p_2,-\hbar \underline{k}_1))\delta^+(2 p_1 \cdot \underline{k}_1)\delta^+(-2 p_2 \cdot \underline{k}_1)\Big)+\mathcal O(\hbar)
\end{multline}
Here, crucially, $N(\gamma,\hbar^2 (\underline{k}_1+\underline{q})^2)$ by construction has only classical and quantum parts. This means that for classical observables $O$ the
matrix element
$A_1^O$ will have a leading piece which is classical, followed by quantum corrections. There are no superclassical pieces in $A_1^O$.
By recursion it follows that this holds for $A_n^O$ and any $n$ as well. 

Although the completeness relation has a positive energy constraint, this is automatically satisfied in
the classical limit for the massive scalars of positive energy,
\begin{equation}\label{e:deltaplusclassical}\delta^+((p_1-\hbar
\underline{k}_1)^2-m_1^2)=\theta(p_1^0-\hbar
\underline{k}_1^0) \delta((p_1-\hbar \underline{k}_1)^2-m_1^2)\simeq
\theta(p_1^0) \delta(-2 \hbar  p_1\cdot
\underline{k}_1)^2) .
\end{equation}
To conclude, we have shown that the classical piece of $A_1^O$ is given by 
\begin{multline}\label{e:A1Oclassical}
  A_1^O(p_1,p_2,q) = \cr
\int \!\!\frac{d^D k_1}{(2\pi)^{D-2}} N(\gamma,(k_1\!+\!q)^2) (k_1^\mu\!+\!q^\mu)\nabla_\mu [O(p_1,p_2,\!- k_1))\delta(2 p_1\! \cdot\! k_1)\delta(-2 p_2\! \cdot\! k_1)]
\end{multline}
after setting $\hbar =1$. Not that this is an all-order statement in $G$. Iterating, it follows that all higher commutators and hence also the full expectation value are free of 
superclassical pieces when
evaluated in the conservative sector.

\subsubsection{Vector operators}\label{sec:vectorcons}

Let us now consider the application of the general iterative
formula of eq.~\eqref{e:A1Oclassical} to a special class of
four-vector operators $ O^\mu(p_1,p_2,q)=\langle p_1',p_2'| \hat O^\mu
|p_1,p_2\rangle$ that decompose into longitudinal $O_{\parallel}(\gamma,q^2)$ and
transverse $O_{\perp}(\gamma,q^2)$ parts as follows:
\begin{equation}\label{e:Odec}
  O^\nu(p_1,p_2,q) = O_{\parallel}((p_1+p_2)^2,q^2)L^{\nu}  + O_{\perp}((p_1+p_2)^2,q^2) q^\nu.
\end{equation}
It is convenient to introduce the four-vector
\begin{equation}\label{e:LmuDef}
L^{\mu} ~\equiv~ \frac{(m_2^2+m_1 m_2
    \gamma)p_1^{\mu} - (m_1^2+m_1 m_2 \gamma)p_2^{\mu}}{m_1^2
    m_2^2(\gamma^2-1)}
\end{equation}
which satisfies nice relations,
\begin{equation}
L\cdot p_2 = 1~,\quad L\cdot p_1 = -1~,\quad b\cdot L=0~, \quad \nabla^{\mu}L_{\mu} ~=~ \frac{1}{p_{\infty}^2}.
\end{equation}
where we used that impact parameter $b^\mu$ lies in the plane
of scattering and is orthogonal to both $p_1^{\mu}$ and $p_2^{\mu}$.
Because $L\cdot q=O(q^2) $, we also have $L\cdot q=0$ in $q$-space, before the Fourier transform to $b$-space. 
Since  $-p_1\cdot q=p_2\cdot q={q^2\over2}$,
$p_1$ and $p_2$ are indeed orthogonal to $q$ in the classical limit. Here,
\beq
p_\infty=\frac{m_1m_2\sqrt{\gamma^2-1}}{\sqrt{m_1^2+m_2^2+2m_1m_2\gamma}}
\eeq

The decomposition in~\eqref{e:Odec} is clearly not valid for an arbitrary four-vector but it is satisfied by the 
momentum kick $\langle\Delta
P_1^\mu\rangle$ when evaluated in the
conservative sector as we will do in section~\ref{sec:DeltaP1cons}. 

To evaluate the classical part of the first commutator $A_1^{O^{\nu}} ={1\over \hbar} \langle p_1',p_2'|[\hat
N,\hat O^{\nu}]| p_1,p_2\rangle$ using the expression~\eqref{e:A1Oclassical}  we begin by acting with the derivative
$\nabla_{\mu}$ in~\eqref{e:NablaF}. It is useful to note that $\nabla_\mu (p_1+p_2)^2=0$ and $\nabla_\mu k_1^\nu=0$
so that  
\beq
\nabla_\mu O_{r}((p_1+p_2)^2,-k_1)=0 
\eeq
for both the longitudinal part $r=\parallel$ and the transverse part $r= \perp$.
We then get 
\begin{multline}
  \nabla_\mu \Big(O^{\nu}(p_1,p_2,- k_1))\delta(2 p_1 \cdot
  k_1)\delta(-2 p_2 \cdot k_1)\Big) =\frac{1}{p_{\infty}^2}
  O_{\parallel}((p_1+p_2)^2,k_1^2) \delta_{\mu}^{ \nu}\delta(2 p_1
  \cdot k_1)\delta(-2 p_2 \cdot k_1) \cr + 
2 k_{1\mu} \Big(O_{\parallel}((p_1+p_2)^2,k_1^2)L^{\nu} - O_{\perp}((p_1+p_2)^2,k_1^2) k_1^{\nu}\Big) \Big(\delta'(2 p_1 \cdot k_1)\delta(-2 p_2 \cdot k_1)+\delta(2 p_1 \cdot k_1)\delta'(-2 p_2 \cdot k_1)\Big) ~.
\end{multline}
which we can insert into eq.~\eqref{e:A1Oclassical}, keeping only the classical
part: 
\begin{multline}
A_1^{O^\nu}(p_1,p_2,q)=\frac{1}{p_{\infty}^2} \int \frac{d^D k_1}{(2\pi)^{D-2}} N(\gamma,(k_1+q)^2) (k_1^\nu+q^\nu)O_{\parallel}((p_1+p_2)^2,k_1^2)\delta(2 p_1 \cdot k_1)\delta(-2 p_2 \cdot k_1) \\+
2\int \frac{d^D k_1}{(2\pi)^{D-2}} N(\gamma,(k_1+q)^2) (k_1+q)\cdot k_1 \Big(O_{\parallel}((p_1+p_2)^2,k_1^2)L^{\nu} \Big)\\ \times \Big(\delta'(2 p_1 \cdot k_1)\delta(-2 p_2 \cdot k_1)+\delta(2 p_1 \cdot k_1)\delta'(-2 p_2 \cdot k_1)\Big)\\
-2\int \frac{d^D k_1}{(2\pi)^{D-2}} N(\gamma,(k_1+q)^2) (k_1+q)\cdot
k_1\Big(O_{\perp}(\gamma,k_1^2) k_1^\nu\Big)\cr
\times\Big(\delta'(2 p_1 \cdot k_1)\delta(-2 p_2 \cdot k_1)+\delta(2 p_1 \cdot k_1)\delta'(-2 p_2 \cdot k_1)\Big).
\end{multline}
By symmetry the integral in the second line vanishes. We thus have
\begin{equation}
A_1^{O^\mu}(p_1,p_2,q) = \frac{1}{p_{\infty}^2} A_1^{O_\parallel \mu }(\gamma,q^2) + A_1^{O_\perp \mu}(\gamma,q^2)\,
\end{equation}
with 
\begin{equation}\label{e:A1par}
 A_1^{O_\parallel \mu}(\gamma,q^2) \equiv  \int \frac{d^D k_1}{(2\pi)^{D-2}} N(\gamma,(k_1+q)^2) (k_1^\mu+q^\mu)O_{\parallel}((p_1+p_2)^2,k_1^2)\delta(2 p_1 \cdot k_1)\delta(-2 p_2 \cdot k_1),
\end{equation}
and
\begin{multline}
  \label{e:A1perp}
  A_1^{O_\perp \mu}(\gamma,q^2)\equiv-2\int \frac{d^D k_1}{(2\pi)^{D-2}} N(\gamma,(k_1+q)^2) (k_1+q)\cdot
k_1\Big(O_{\perp}(\gamma,k_1^2) k_1^\mu\Big)\cr
\times\Big(\delta'(2 p_1 \cdot k_1)\delta(-2 p_2 \cdot k_1)+\delta(2 p_1 \cdot k_1)\delta'(-2 p_2 \cdot k_1)\Big).
\end{multline}
By tensor reduction the latter takes the form

\begin{equation}
  \label{e:A1perp2}
  A_1^{O_\perp \mu}(\gamma,q^2)=- L^\mu \int \frac{d^D k_1}{(2\pi)^{D-2}} N(\gamma,(k_1+q)^2) (k_1+q)\cdot
k_1 O_{\perp}(\gamma,k_1^2) \delta(2 p_1 \cdot k_1)\delta(-2 p_2 \cdot k_1).
\end{equation}

We note an interesting swap between longitudinal and transverse parts in this first iteration. Clearly, when we iterate further, this will generate 
alternating contributions between the longitudinal and transverse parts.

To complete the evaluation of the observable according to the KMOC prescription we now perform the  Fourier transform to $b$-space
according to eq.~\eqref{KMOCbspace}. Having already taken the classical limit, it is clear that we can also ignore the $q^2$-terms
in the two delta-functions and effectively the Fourier transform simply becomes
\begin{equation}\label{e:Fourier}
\tilde O(\gamma,b)=\int \frac{d^D q}{(2\pi)^{D-2}}\delta(-2
p_1 \cdot q)\delta(2 p_2 \cdot q) O((p_1+p_2)^2,q^2)e^{i b \cdot q}.
\end{equation}
For the longitudinal part we have to evaluate the Fourier transform of
$A_1^{O_{\parallel} \mu}(\gamma,q^2)$ which reads
\begin{multline}\label{e:FL1}
\int \frac{d^D q}{(2\pi)^{D-2}} \frac{d^D k_1}{(2\pi)^{D-2}}
N(\gamma,(k_1+q)^2) (q^{\mu}+k_1^{\mu}) O_{\parallel}((p_1+p_2)^2,k_1^2) \delta(2 p_1 \cdot k_1)\delta(-2 p_2 \cdot k_1)\cr\times\delta(-2 p_1 \cdot q)\delta(2 p_2 \cdot q)e^{i b \cdot q} .
\end{multline}
and by a change of variables $q \rightarrow q-k_1$ and $k_1
\rightarrow -k_1$ the integral  factorizes
\begin{multline}
\eqref{e:FL1}=\int \frac{d^D q}{(2\pi)^{D-2}} q^\nu
N(\gamma,q^2) 
\delta(-2 p_1 \cdot q)\delta(2 p_2 \cdot q)e^{i b \cdot
  q}
\cr
\times
\int 
\frac{d^D k_1}{(2\pi)^{D-2}} 
O_{\parallel}((p_1+p_2)^2,k_1^2) \delta(-2 p_1 \cdot k_1)\delta(2 p_2 \cdot
k_1)\,e^{i b \cdot k_1}.
\end{multline}
Setting
\begin{equation}
  \tilde{O}_{\parallel}(\gamma,b)\equiv\int  \frac{d^D k_1}{(2\pi)^{D-2}} 
O_{\parallel}((p_1+p_2)^2,k_1^2) \delta(-2 p_1 \cdot k_1)\delta(2 p_2 \cdot
k_1)\,e^{i b \cdot k_1} 
\end{equation}
and noticing that
\begin{equation}
-i{\partial \tilde N(\gamma,b)\over \partial b_\nu}=\int \frac{d^D q}{(2\pi)^{D-2}} q^\nu
N(\gamma,q^2) 
\delta(-2 p_1 \cdot q)\delta(2 p_2 \cdot q)e^{i b \cdot
  q}.
\end{equation}
with $\tilde{N}(\gamma,J) $ the Fourier transform of $N(\gamma,q^2)$
to $b$-space
\begin{equation}
\tilde{N}(\gamma,b) ~\equiv~  \text{FT}[N(\gamma,q^2)]~\equiv~ \frac{1}{4m_1m_2\sqrt{\gamma^2 - 1}}\int \frac{d^2q}{(2\pi)^2} N(\gamma,q^2) e^{i{b}\cdot{q}}.
\label{FTdef}
\end{equation}
we find that the Fourier transform of
$A_1^{O_{\parallel} \mu}(\gamma,q^2)$  is given by
\begin{equation}
  - i \frac{\partial
      \tilde{N}(\gamma,b)}{\partial b_{\nu}} \tilde{O}_{\parallel}(\gamma,b) =
 i {b^\nu\over |b|} \frac{\partial
      \tilde{N}(\gamma,b)}{\partial |b|} \tilde{O}_{\parallel}(\gamma,b) .
\end{equation}
For  the transverse part we have to evaluate
\begin{multline}
\tilde{A}_1^{O_\perp \mu}(\gamma,b)=- L^\mu  \int \frac{d^D q}{(2\pi)^{D-2}}
\frac{d^D k_1}{(2\pi)^{D-2}} N(\gamma,(k_1+q)^2) (k_1+q)\cdot
k_1 O_{\perp}(\gamma,k_1^2) \cr \times \delta(2 p_1 \cdot k_1)\delta(-2 p_2 \cdot k_1)\delta(-2 p_1 \cdot q)\delta(2 p_2 \cdot q)e^{i b \cdot q}
\end{multline}
By the same change of variable as before we get

\begin{multline}
\tilde{A}_1^{O_\perp \mu}(\gamma,b)=L^\mu  \int \frac{d^D q}{(2\pi)^{D-2}}
\frac{d^D k_1}{(2\pi)^{D-2}} N(\gamma,q^2) q\cdot
k_1 O_{\perp}(\gamma,k_1^2) \cr \times \delta(-2 p_1 \cdot k_1)\delta(2 p_2 \cdot k_1)\delta(-2 p_1 \cdot q)\delta(2 p_2 \cdot q)e^{i b \cdot q}e^{i b \cdot k_1}
\end{multline}

This integral is product of a Fourier transform over $q$ times a
Fourier transform over $k_1$ 
leading to 
\begin{equation}
\tilde{A}_1^{O_\perp \mu}(\gamma,b)=-L^\mu \frac{\partial \tilde{N}(\gamma,b)}{\partial
  b^\nu} \frac{\partial \tilde{O}_{\perp}(\gamma,b)}{\partial b_\nu}=L^\mu  \frac{\partial \tilde{N}(\gamma,b)}{\partial
  |b|} \frac{\partial \tilde{O}_{\perp}(\gamma,b)}{\partial |b|}.
\end{equation}
Collecting these pieces, we get
\begin{equation}
\tilde{A}_1^{O^\mu}(\gamma,b)=\Bigg( \frac{i}{p_{\infty}^2} \frac{b^\nu}{\vert b \vert} \tilde{O}_{\parallel}(\gamma,b) + L^{\nu} 
\frac{\partial \tilde{O}_{\perp}(\gamma,b)}{\partial \vert b \vert}\Bigg)\frac{\partial \tilde{N}(\gamma,b)}{\partial \vert b \vert}.
\end{equation}
In term of the angular momentum $J=p_\infty |b|$, we have
\begin{equation}\label{e:A10}
\tilde{A}_1^{O^\mu}(\gamma,b)=\Bigg( \frac{i}{p_{\infty}} \frac{b^\mu}{\vert b \vert} \tilde{O}_{\parallel}(\gamma,b) +p_\infty L^{\mu} 
\frac{\partial \tilde{O}_{\perp}(\gamma,b)}{\partial \vert b \vert}\Bigg)\frac{\partial \tilde{N}(\gamma,J)}{\partial \vert J \vert}.
\end{equation}
The factorization of the Fourier transforms separates the $N$ operator
from the operator $O$ in $b$-space.
This remarkable fact implies that we can iterate the result above as
dictated by the commutator relation in eq.~\eqref{e:AOn}. It is convenient to introduce a
matrix notation so that

\begin{equation}
    \tilde{A}_{1}^{O^\mu}(\gamma,b)=\begin{pmatrix}L^{\mu}& 
i      \frac{b^\mu}{|b|}\end{pmatrix}
     \begin{pmatrix}
0 & p_{\infty}\frac{\partial \tilde{N}}{\partial \vert J \vert} \\
\frac{1}{p_{\infty}} \frac{\partial \tilde{N}}{\partial \vert J \vert} &0
\end{pmatrix}
\begin{pmatrix}
\tilde O_\parallel \cr \partial\tilde O_\perp \over \partial|b|
\end{pmatrix}
\end{equation}
and

\begin{equation}
    \tilde{A}_{n+1}^{O^\mu}(\gamma,b)=\begin{pmatrix}L^{\mu}& 
i      \frac{b^\mu}{|b|}\end{pmatrix}
     \begin{pmatrix}
0 & p_{\infty}\frac{\partial \tilde{N}}{\partial \vert J \vert} \\
\frac{1}{p_{\infty}} \frac{\partial \tilde{N}}{\partial \vert J \vert} &0
\end{pmatrix}^n
\begin{pmatrix}
{\partial\tilde O_\perp \over \partial|b|} {\partial\tilde
  N\over \partial J} \cr  {\tilde O_\parallel\over p_\infty}  {\partial\tilde N\over \partial J}
\end{pmatrix}
\end{equation}
for summing the iteration to all orders according the recursion in
eq.~\eqref{e:AOn}.
Inserting it in the expression~\eqref{e:DeltaOcomm}, we get

\begin{equation}
\Delta \tilde{O}^{\mu}(\gamma,b)=\begin{pmatrix}L^{\mu}& 
   i  \frac{b^\mu}{|b|}\end{pmatrix}
\sum_{n =1}^{\infty}  \frac{(-i)^n}{n!} \begin{pmatrix}
0 & p_{\infty}\frac{\partial \tilde{N}}{\partial \vert J \vert} \\
\frac{1}{p_{\infty}} \frac{\partial \tilde{N}}{\partial \vert J \vert} &0
\end{pmatrix}^{n-1}\begin{pmatrix}
p_\infty {\partial\tilde O_\perp \over \partial|b|} {\partial\tilde
  N\over \partial J}\cr  {\tilde O_\parallel\over p_\infty}  {\partial\tilde N\over \partial J}
\end{pmatrix}.
\end{equation}
which sums into
\begin{equation}\label{e:DeltaOconsFinal}
\Delta \tilde{O}^{\mu}(\gamma,b)=
\begin{pmatrix}L^{\mu}& 
    i\frac{b^\mu}{|b|}\end{pmatrix}
\begin{pmatrix}
 -\frac{i \sin\left(\partial \tilde N\over \partial J\right)}{{\partial\tilde N\over\partial J}} &
   \frac{p_\infty (\cos
  \left(\partial \tilde N\over \partial J\right)-1)}{{\partial\tilde N\over\partial J}} \\
 \frac{\cos\left(\partial \tilde N\over \partial J\right)-1}{{\partial\tilde N\over\partial J} p_\infty   } & -\frac{i \sin\left(\partial \tilde N\over \partial J\right)}{{\partial\tilde N\over\partial J}}
\end{pmatrix}
\begin{pmatrix}
p_\infty {\partial\tilde O_\perp \over \partial|b|} {\partial\tilde
  N\over \partial J}\cr  {\tilde O_\parallel\over p_\infty}  {\partial\tilde N\over \partial J}
\end{pmatrix}.
\end{equation}
This relation shows the intimate connection between the exponential representation of the $S$-matrix and the KMOC formalism. It is an interesting fact that the $\hat{N}$-operator is
here sandwiched between the initial {\em in}-state and its conjugate rather than between {\em in} and {\em out} states as in ref. \cite{Damgaard:2021ipf}. This is
a consequence of the fact that the KMOC formalism evaluates observables as the difference between time evolved {\em in}-states whereas in \cite{Damgaard:2021ipf} 
$N(\gamma,b)$ was viewed as an ordinary scattering matrix element from which to compute the scattering angle through the radial action. It is also interesting to note how the iterative structure of the exponential
representation makes $\hat{N}$ matrix elements the universal objects to compute in the KMOC formalism, whereas all details of the actual observable $O^\mu$ only enter through the
initial vector determined by $\tilde{A}_1^{O^\nu}(\gamma,b)$ in~\eqref{e:A10}.

\subsubsection{Momentum kick: the conservative sector}\label{sec:DeltaP1cons}

We now finally apply the general considerations above to the case of
the momentum kick of, say, particle 1 with initial  momentum $p_1$ in
the scattering. We then have that the initial vector is 
\begin{equation}
\tilde  A^{P_1^\mu}(\gamma,b)= i p_\infty {b^\mu\over|b|} {\partial
  \tilde N(\gamma,J)\over \partial J}.
\end{equation}
We apply the equation~\eqref{e:DeltaOconsFinal} with $\tilde
O_\parallel(\gamma,b)=p_\infty^2$ and $\tilde O_\perp(\gamma,b)=0$. We get
\begin{equation}
\Delta \tilde{P_1}^\nu(\gamma,b)|_{\rm cons}=p_\infty \frac{b^\nu}{\vert b \vert} \sin \bigg(\frac{\partial \tilde{N}(\gamma,J)}{\partial J}\Bigg) \\
+ p_\infty^2 L^{\nu} \Bigg(\cos \Bigg(-\frac{\partial \tilde{N}(\gamma,J)}{\partial J}\Bigg)-1\Bigg).
\label{ConsDelta}
\end{equation}
In the conservative case, the scattering angle can be extracted by the coefficient of the transverse piece only.
A comparison with the general relation between momentum kick and scattering angle~\cite{Herrmann:2021tct}\footnote{The coefficient of $\sin(\chi)$ is fixed
by a quadratic condition. We choose the sign opposite to that of ref.~\cite{Herrmann:2021tct}.}
\begin{equation}
\Delta \tilde{P_1}^\nu(\gamma,b)|_{\rm cons}=-p_\infty \frac{b^\nu}{\vert b \vert} \sin(\chi) \\+ p_\infty^2 L^{\nu} \left(\cos(\chi)-1\right),
\end{equation}
demonstrates that
\begin{equation}\label{e:chitoN}
\chi ~=~ -{\partial \tilde N(\gamma,J)\over \partial J}~=~ - \frac{1}{p_{\infty}} \frac{\partial \tilde{N}(\gamma,b)}{\partial b} ~,
\end{equation}
thus proving the conjectured relation of ref.~\cite{Damgaard:2021ipf}
between the scattering angle and the matrix elements of the
$N$-operator. This also shows that the $\tilde N(\gamma,J)$ is the radial action.

\subsection{Including gravitational radiation}\label{sec:radiation}
We now turn to the impact of gravitational radiation on the expectation
value of an operator $\hat O$. We recall that in the KMOC formalism radiation is automatically taken into account in perturbation
theory by insertion of a complete set of states (including any number of gravitons) in the pertinent {\em in-in} matrix elements.
Conventionally done by means of the Born expansion of the $\hat T$-matrix, we here adapt it to the exponential representation.
In particular, we use the insertion of the identity operator inside the nested commutators and extract contributions order
by order in the gravitational coupling $G$. To clarify: when going
from $\hat T$-matrix elements to $\hat N$-matrix elements we also include 
terms that are radiative,
to arbitrarily high order in the coupling $G$.
What is missing in order to compute the full expectation value of an operator $\hat O$ are the pieces that arise from 
inserting complete sets of states 
(including gravitons) {\em inside the nested commutators of eq. (\ref{e:DeltaOcomm})}. The discussion will clearly mimic closely the
way we evaluated matrix elements of $\hat{N}$-operator itself. We now consider these additional terms.

Since our aim is to derive a recursive relation for the classical limit of an observable, 
we begin by analyzing the expectation value of $\hat
A^{\hat{O}}_{n+1}$ based on one iteration,
\begin{equation}
\langle p_1',p_2'|
    \hat{A}^{\hat{O}^\mu}_{n+1}|p_1,p_2\rangle={1\over\hbar}\langle p_1',p_2'|[\hat N,  \hat{A}^{\hat{O}^\mu}_{n}]|p_1,p_2\rangle.
  \end{equation}
Inserting a complete set of states, this has a graphical representation
\begin{multline}\label{e:A20consrad}
 A_{n+1}^{O^\mu}(p_1,p_2,q) =
 \begin{gathered}
\begin{tikzpicture}[scale=0.5]
\draw [black,very thick] [->] (-4,-.8) to (4,-.8);
\draw [black,very thick] [->] (-4,.8) to (4,.8);
\draw [red] (0,-1) to (0,-.6);
\draw [red] (0,1) to (0,.6);
\filldraw [color = black, fill=white, very thick] (-2,0) circle (1cm)
node {$N$};
\filldraw [color = black, fill=white, very thick] (2,0) circle (1cm)
node {$A_n^{O^\mu}$};
\end{tikzpicture}
\end{gathered}
-
 \begin{gathered}
\begin{tikzpicture}[scale=0.5]
\draw [black,very thick] [->] (-4,-.8) to (4,-.8);
\draw [black,very thick] [->] (-4,.8) to (4,.8);
\draw [red] (0,-1) to (0,-.6);
\draw [red] (0,1) to (0,.6);
\filldraw [color = black, fill=white, very thick] (-2,0) circle (1cm)
node {$A_n^{O^\mu}$};
\filldraw [color = black, fill=white, very thick] (2,0) circle (1cm)
node {$N$};
\end{tikzpicture}
\end{gathered}\cr
+
\begin{gathered}
\begin{tikzpicture}[scale=0.5]
\draw [black,very thick] [->] (-4,-.8) to (4,-.8);
\draw [black,very thick] [->] (-4,.8) to (4,.8);
\draw [black,snake it]  (-1.5,0) to (1.5,0);
\draw [red] (0,-1) to (0,-.6);
\draw [red] (0,1) to (0,.6);
\draw [red] (0,.3) to (0,-.3);
\filldraw [color = black, fill=white, very thick] (-2,0) circle (1cm)
node {$N$};
\filldraw [color = black, fill=white, very thick] (2,0) circle (1cm)
node {$A_n^{O^\mu}$};
\end{tikzpicture}
\end{gathered}
-
 \begin{gathered}
\begin{tikzpicture}[scale=0.5]
\draw [black,very thick] [->] (-4,-.8) to (4,-.8);
\draw [black,very thick] [->] (-4,.8) to (4,.8);
\draw [black,snake it]  (-1.5,0) to (1.5,0);
\draw [red] (0,-1) to (0,-.6);
\draw [red] (0,1) to (0,.6);
\draw [red] (0,.3) to (0,-.3);
\filldraw [color = black, fill=white, very thick] (-2,0) circle (1cm)
node {$A_n^{O^\mu}$};
\filldraw [color = black, fill=white, very thick] (2,0) circle (1cm)
node {$N$};
\end{tikzpicture}
\end{gathered}+\cdots
\end{multline}
where the ellipsis represent pieces with insertion of more that one graviton. We stress that this involves the full
$\hat{N}$-operator and in perturbation theory we obviously need to truncate to the given order in $G$ (but for now
we keep it general). Up to $\mathcal O(G^4)$ we only need to consider the $n=2$ term and to compute the iteration of this term, i.e. the $n=3$ contribution with one graviton insertion

\begin{equation}
[\hat{N}^{\text{rad}},[\hat{N}^{\text{rad}},[\hat{N},\hat{O}]]]+[\hat{N}^{\text{rad}},[\hat{N},[\hat{N}^{\text{rad}},\hat{O}]]]+[\hat{N},[\hat{N}^{\text{rad}},[\hat{N}^{\text{rad}},\hat{O}]]] ~.
\end{equation}
By simple manipulations this can be written

\beq
[[ \hat{N}^{\text{rad}},[ \hat{N}^{\text{rad}},\hat{N}]],\hat{O}] + 3 [\hat{N},[ \hat{N}^{\text{rad}}, [\hat{N}^{\text{rad}},\hat{O}]]] + 3[[\hat{N}^{\text{rad}},\hat{N}],[\hat{N}^{\text{rad}},\hat{O}]]
~.
\eeq
Taking the classical limit, we find that the first and last terms vanish so that we are left with
$$
3 [\hat{N},[ \hat{N}^{\text{rad}}, [\hat{N}^{\text{rad}},\hat{O}]]]
$$
This term can be evaluated using the same tools we developed in the previous part for the conservative pieces.
 
 This concludes the analysis of single-graviton insertions from the complete set of states up to $\mathcal O(G^4)$. Actually, 
what we just shown can be generalized to any number of graviton insertions. However, at three-loop level, and as noticed in ref.~\cite{Cristofoli:2021jas}
in the context of the eikonal, multiple graviton insertions such as

\begin{equation}
\begin{gathered}
\begin{tikzpicture}[scale=0.5]
\draw [black,very thick] [->] (-4,-.8) to (4,-.8);
\draw [black,very thick] [->] (-4,.8) to (4,.8);
\draw [black,snake it]  (-1.5,0.3) to (1.5,0.3);
\draw [black,snake it]  (-1.5,-0.3) to (1.5,-0.3);
\draw [red] (0,-1) to (0,-.6);
\draw [red] (0,1) to (0,.6);
\draw [red] (0,.15) to (0,.45);
\draw [red] (0,-.15) to (0,-.45);
\filldraw [color = black, fill=white, very thick] (-2,0) circle (1cm)
node {};
\filldraw [color = black, fill=white, very thick] (2,0) circle (1cm)
node {};
\end{tikzpicture}
\end{gathered}
\end{equation}
do not contribute classically. To fourth Post-Minkowskian order we thus only need to consider successions of single gravitons insertions such as
 
 \begin{equation}
\begin{gathered}
\begin{tikzpicture}[scale=0.5]
\draw [black,very thick] [->] (-8,-.8) to (4,-.8);
\draw [black,very thick] [->] (-8,.8) to (4,.8);
\draw [black,snake it]  (-6,0) to (-2,0);
\draw [black,snake it]  (-2,0) to (2,0);
\draw [red] (-4,-1) to (-4,-.6);
\draw [red] (-4,1) to (-4,.6);
\draw [red] (0,1) to (0,.6);
\draw [red] (0,-1) to (0,-.6);
\draw [red] (-4,.3) to (-4,-.3);
\draw [red] (0,.3) to (0,-.3);
\filldraw [color = black, fill=white, very thick] (-6,0) circle (1cm)
node {};
\filldraw [color = black, fill=white, very thick] (-2,0) circle (1cm)
node {};
\filldraw [color = black, fill=white, very thick] (2,0) circle (1cm)
node {};
\end{tikzpicture}
\end{gathered}
\end{equation}
 
When we include these radiative pieces we need to enlarge the basis for vector operators. We choose to introduce 
\begin{equation}
u_1^\mu \equiv p_\infty \frac{m_1 \gamma p_2^\mu-m_2 p_1^\mu}{m_1^2
    m_2(\gamma^2-1)}, \qquad u_2^\mu \equiv p_\infty \frac{m_2 \gamma p_1^\mu-m_1
    p_2^\mu}{m_1 m_2^2(\gamma^2-1)},
\end{equation}
which satisfy $p_i\cdot u_j=p_\infty \delta_{ij}$.
These two four-vectors are  related to the $\check{u}_i$'s
of~\cite{Herrmann:2021tct} by a rescaling
$u_i=\check u_i p_\infty/m_i$ with $i=1,2$.
The vector $L^\mu$ of eq.~\eqref{e:LmuDef} which sufficed to describe the basis in the conservative
sector is simply a specific linear combination,
\begin{equation}\label{e:Lsplit}
  L^\mu= (u_2^\mu-u_1^\mu)/ p_\infty  .
\end{equation}
We now need to consider a vector $X^\mu$ that can be decomposed according to

\begin{equation}
X^\mu=\frac{p_1.X}{p_\infty}u_1^\mu+\frac{p_2.X}{p_\infty}u_2^\mu+ \Big(-\frac{b.X}{|b|}\Big)\frac{b^{\mu}}{|b|}\equiv X^{u_1}u_1^\mu+X^{u_2}u_2^\mu+X^{b}\frac{b^{\mu}}{|b|}
\end{equation}
Manipulations similar to the ones of 
section~\ref{sec:conservative} yield a compact matrix identity 
in $b$-space, now taking
into account the radiation effects with at most one graviton exchange. For the conservative insertions we have
\begin{equation}
  \langle p_1',p_2'|\hat{A}^{\hat{O}^\mu}_{n+1}|p_1,p_2\rangle=
  \begin{pmatrix}
    u_1^\mu& u_2^\mu&{b^\mu\over |b|}
  \end{pmatrix}
  M
\begin{pmatrix}
\langle p_1',p_2'|\hat{A}^{\hat{O}}_{n}|p_1,p_2\rangle^{u_1} \cr  \langle p_1',p_2'|\hat{A}^{\hat{O}}_{n}|p_1,p_2\rangle^{u_2} \cr \langle p_1',p_2'|\hat{A}^{\hat{O}}_{n}|p_1,p_2\rangle^{b} \end{pmatrix} 
\end{equation}
where we have defined the matrix
\begin{equation}
 M\equiv  
  \begin{pmatrix}
0 &0&i{\partial \tilde N\over \partial J}\\
0 &0&-i{\partial \tilde N\over \partial J}\\-
{i \mathcal E_2} {\partial \tilde N\over \partial J}&{i\mathcal E_1}{\partial \tilde N\over \partial J}&0
\end{pmatrix}
\end{equation}
and we have introduced fractional parts $\mathcal E_1$ and $\mathcal E_2$  of the Mendelstam variable $s$ by definitions 

\begin{equation}\label{e:E1E2def}
  \mathcal
  E_1\equiv {m_1^2+m_1m_2\gamma\over m_1^2+m_2^2+2m_1m_2\gamma}; \qquad
  \mathcal E_2\equiv 1-\mathcal E_1= {m_2^2+m_1m_2\gamma\over m_1^2+m_2^2+2m_1m_2\gamma}.
\end{equation}
As we have shown, at fourth Post-Minkowskian order we can write the full result as 
\begin{equation}
  \Delta \tilde{O}(\gamma,b)= \Delta \tilde{O}_{{\rm cons}}(\gamma,b )+\sum_{n=1}^{\infty}{\Delta \tilde{O}_{{\rm rad}}^{(n)}(\gamma,b)}
\end{equation}
where $\Delta \tilde{O}_{{\rm rad}}^{(n)}$ is the contribution coming from the succession of $n$ single-graviton insertions. The conservative part is given by
\begin{align}
&    \Delta \tilde{O}_{{\rm cons}}(\gamma,b )=\begin{pmatrix}
    u_1^\mu&  u_2^\mu& {b^\mu\over |b|}
  \end{pmatrix}\sum_{n \geq 1}  \frac{(-i)^n}{n!}
M^{n-1} \begin{pmatrix}
  {\tilde O^{u_1}_1} \cr
{\tilde O^{u_2}_1} \cr
 {\tilde O^{b}_1}
\end{pmatrix}\\
\nonumber &=i\begin{pmatrix}
    u_1^\mu& u_2^\mu & {b^\mu\over |b|}
  \end{pmatrix} \begin{pmatrix}
  -\frac{({\partial\tilde N\over\partial J} \mathcal E_1+\mathcal E_2 \sin
   \left(\partial \tilde N\over \partial J\right))}{{\partial\tilde N\over\partial J}} & -\frac{ \mathcal E_1
   ({\partial\tilde N\over\partial J}-\sin \left(\partial \tilde N\over \partial J\right))}{{\partial\tilde N\over\partial J}} &
   \frac{-1+ \cos
   \left(\partial \tilde N\over \partial J\right)}{{\partial\tilde N\over\partial J} } \\
 -\frac{ \mathcal E_2 ({\partial\tilde N\over\partial J}-\sin
   \left(\partial \tilde N\over \partial J\right))}{{\partial\tilde N\over\partial J}} & -\frac{(\mathcal E_1
   \sin \left(\partial \tilde N\over \partial J\right)+{\partial\tilde N\over\partial J}
   \mathcal E_2)}{{\partial\tilde N\over\partial J}} & \frac{1-\cos \left(\partial \tilde N\over \partial J\right)}{{\partial\tilde N\over\partial J}} \\
 \frac{\mathcal E_2\left(1-\cos
   \left(\partial \tilde N\over \partial J\right)\right)}{{\partial\tilde N\over\partial J}
 } &
   \frac{\mathcal E_1 (\cos \left(\partial \tilde N\over \partial J\right)-1)}{{\partial\tilde N\over\partial J}
   } & -\frac{\sin
   \left(\partial \tilde N\over \partial J\right)}{{\partial\tilde N\over\partial J}}
\end{pmatrix} \begin{pmatrix}
  {\tilde O^{u_1}_1} \cr
{\tilde O^{u_2}_1} \cr
 {\tilde O^{b}_1}
\end{pmatrix},
\end{align}
where we have introduced the  operator $ \hat O_1\equiv  [\hat N, \hat
O]$.  This is just a different way of writing the conservative result
of eq.~\eqref{e:DeltaOconsFinal}, as can be seen by use of the
relations~\eqref{e:Lsplit} and~\eqref{e:E1E2def}. 

For the radiative sector we get a similar formula

\begin{equation}
\Delta \tilde{O}_{{\rm rad}}^{(1)}(\gamma,b)=\begin{pmatrix}
    u_1^\mu & u_2^\mu & {b^\mu\over |b|}
  \end{pmatrix} \Bigg( -\frac{1}{2}+ \frac{i M}{2} \Bigg) \begin{pmatrix}
  {\tilde O^{u_1}_{2}} \cr
{\tilde O^{u_2}_{2}} \cr
 {\tilde O^{b}_{2}}
\end{pmatrix}+\mathcal O(G^5)
\end{equation}
where we defined 
\beq
\hat{O}_{k+1} = \underbrace{[\hat{N},[\hat{N},\dots,[\hat{N},\hat{O}]]]}_{\text{k+1 times}}|_{\text{k graviton insertions}}
\eeq
after restricting to $k$ graviton insertions, as explained above. This is the complete expression to fourth Post-Minkowskian order and it
is readily generalized to higher orders.

We emphasize again that the terminology of conservative and radiative pieces is completely artificial. There are also radiative modes in what we for
historical reasons call the conservative part. This was already obvious at two-loop level where it was shown in refs.~\cite{Damgaard:2021ipf}
that the two-to-two matrix element of $\hat{N}$-operator yields the full result, including radiation reaction, to that order. We now understand why
this phenomenon does not generalize to higher orders, and we understand how to correct for it. There are still many radiative modes and radiation-reaction
parts in just the two-to-two matrix element of $\hat{N}$-operator and therefore those matrix elements are far from being just conservative.

\subsubsection{Full momentum kick at fourth Post-Minkowskian order}\label{sec:P1radiation}

We now turn to the full explicit evaluation of the
momentum kick $\Delta P_1^\mu$ at fourth Post-Minkowskian order. As a building block we will first need to compute $\tilde{N}(\gamma,b)$. This was 
already done in ref.~\cite{Bjerrum-Bohr:2022ows} up to 4PM order (except for one term which we take the opportunity to correct here) so that what we
label the conservative piece

\begin{equation}
\Delta P_1^\mu|_{{\rm cons.}}=\begin{pmatrix}
    u_1^\mu & u_2^\mu & {b^\mu\over |b|}
  \end{pmatrix}\begin{pmatrix}
  {p_\infty \Big(1- \cos(\chi_{\text{cons}})\Big)} \cr
{p_\infty \Big(\cos(\chi_{\text{cons}})-1\Big)} \cr
 {-p_\infty \sin(\chi_{\text{cons}})}
 \end{pmatrix}
\end{equation}
is known. Here it is convenient to introduce the following notation

\begin{equation}
\chi_{\text{cons}}\equiv-\frac{\partial \tilde N}{\partial J}
\end{equation}
and define the PM-expanded quantities

\begin{equation}
\chi_{\text{cons}}\equiv \sum_{n=0}^{\infty} G^{n+1} \chi^{(n)}_{\text{cons}}
\end{equation}
as well as

\begin{equation}
\tilde N \equiv \sum_{n=0}^{\infty} G^{n+1} \tilde N^{(n)}
\end{equation}
so that at fourth Post-Minkowskian order we have 

\begin{align}
\Delta P_1^{\mu,4PM}|_{{\rm cons.}}&=p_\infty G^4 \begin{pmatrix}
    u_1^\mu & u_2^\mu & {b^\mu\over |b|}
  \end{pmatrix}\begin{pmatrix}
  {- \frac{ (\chi_{\text{cons}}^{(0)})^4}{24}+\frac{(\chi_{\text{cons}}^{(1)})^2}{2}+ \chi_{\text{cons}}^{(0)}\chi_{\text{cons}}^{(2)}} \cr
{\frac{ (\chi_{\text{cons}}^{(0)})^4}{24}-\frac{(\chi_{\text{cons}}^{(1)})^2}{2}- \chi_{\text{cons}}^{(0)}\chi_{\text{cons}}^{(2)}} \cr
{\frac{(\chi_{\text{cons}}^{(0)})^2\chi_{\text{cons}}^{(1)}}{2}-\chi_{\text{cons}}^{(3)}}
 \end{pmatrix}
\end{align}

Starting at third Post-Minkowskian order we need to also evaluate the first radiation contribution to the momentum kick
 $\Delta \tilde{P}_{{1,\rm rad}}^{\mu (1)}$. We thus need the building block 

\begin{equation}
\tilde{P}_{1,1}^\mu=\langle p_1',p_2' | [\hat{N},[\hat{N},\hat{P}_1^\mu]]|p_1,p_2\rangle
\end{equation}
evaluated with one-graviton insertions. This reads

\begin{multline}
\tilde{P}_{1,1}^\mu=\text{FT}[\int \frac{d^D q_1 d^D q_2}{(2\pi)^{2D-4}} \langle p_1',p_2' \vert \hat{N}\vert p_1+q_1,p_2-q_2,q_2-q_1\rangle (-q^\mu-2q_1^\mu)\\ \times  \langle p_1+q_1,p_2-q_2,q_2-q_1 \vert \hat{N}\vert p_1,p_2\rangle \delta(2p_1 \cdot q_1)\delta(-2p_2 \cdot q_2)\delta((q_2-q_1)^2)]
\end{multline}
Where again, for compactness of notation, we label the Fourier transform into $b$-space by FT. Its precise definition is given in
eq.~(\ref{FTdef}). Note that this integral is orthogonal to $p_1$, {\em i.e.} 
\begin{equation}
p_{1\mu} \langle
p_1',p_2'|[\hat{N},[\hat{N},\hat{P}_1^\mu]]|p_1,p_2\rangle =0 , 
\end{equation}
so that it can be decomposed according to

\begin{equation}
\tilde{P}_{1,1}^\mu=\begin{pmatrix}
    u_1^\mu & u_2^\mu & {b^\mu\over |b|}
  \end{pmatrix}\begin{pmatrix}
  0 \cr
\tilde{P}_{1,1}^{u_2}  \cr
\tilde{P}_{1,1}^{b}
 \end{pmatrix}
\end{equation}

Based on the analysis of ref.~\cite{Herrmann:2021tct} we know that the
coefficients have the following perturbative expansion
\begin{align}
\tilde{P}_{1,1}^{u_2} &=G^3 \tilde{P}_{1,1}^{u_2,(2)} +G^4 \tilde{P}_{1,1}^{u_2,(3)} +\mathcal O(G^5), \cr
      \tilde{P}_{1,1}^{b}&=G^4 \tilde{P}_{1,1}^{b,(3)}+\mathcal O(G^5).
\end{align}
so that 
\begin{equation}
\Delta \tilde{P}_{1, \text{rad}}^{\nu,(1)} = G^3 \begin{pmatrix}
    u_1^\mu & u_2^\mu & {b^\mu\over |b|}
  \end{pmatrix}\begin{pmatrix}
0 \\
 -\frac{\tilde{P}_{1,1}^{u_2,(2)}}{2}\\
0
\end{pmatrix}  +G^4 \begin{pmatrix}
    u_1^\mu & u_2^\mu & {b^\mu\over |b|}
  \end{pmatrix} \begin{pmatrix}
0 \\
-\frac{\tilde{P}_{1,1}^{u_2,(3)}}{2} \\
\frac{\mathcal E_1 \chi_{\text{cons}}^{(0)}\tilde{P}_{1,1}^{u_2,(2)}}{2} -\frac{\tilde{P}_{1,1}^{b,(3)}}{2} 
\end{pmatrix}+\mathcal O(G^5)
\end{equation}
Note in particular that $\tilde{P}_{1,1}^{b}$ only receives a contribution from order ${\cal O}(G^4)$, the 4PM order. As mentioned above, the 3PM
case is therefore quite special in that all radiative effects are entirely contained in the classical contribution from the $\hat{N}$-operator \cite{Damgaard:2021ipf}. 
The momentum kick due to radiation at 3PM order only shifts
the longitudinal momenta.

Starting at fourth Post-Minkowskian order we need to also evaluate of the first radiative contribution to the momentum kick 
$\Delta \tilde{P}_{{1,\rm rad}}^{\mu (2)}$ which,
as indicated, involves the insertion of two graviton lines. This contribution is more tricky and is diagrammatically represented by

\begin{equation}
\begin{gathered}
\begin{tikzpicture}[scale=0.5]
\draw [black,very thick] [->] (-8,-.8) to (4,-.8);
\draw [black,very thick] [->] (-8,.8) to (4,.8);
\draw [black,snake it]  (-6,0) to (-2,0);
\draw [black,snake it]  (-2,0) to (2,0);
\draw [red] (-4,-1) to (-4,-.6);
\draw [red] (-4,1) to (-4,.6);
\draw [red] (0,1) to (0,.6);
\draw [red] (0,-1) to (0,-.6);
\draw [red] (-4,.3) to (-4,-.3);
\draw [red] (0,.3) to (0,-.3);
\filldraw [color = black, fill=white, very thick] (-6,0) circle (1cm)
node {};
\filldraw [color = black, fill=white, very thick] (-2,0) circle (1cm)
node {};
\filldraw [color = black, fill=white, very thick] (2,0) circle (1cm)
node {};
\end{tikzpicture}
\end{gathered}
\end{equation}
which has two pieces at 4PM order:

\begin{equation}
\begin{gathered}\vspace{.3cm}
\begin{tikzpicture}[scale=0.5]
\draw [black,very thick] [->] (-8,-.8) to (4,-.8);
\draw [black,very thick] (-8,.8) to (-6,.8);
\draw [black,very thick] [->] (2,.8) to (4,.8);
\draw [black,snake it] [->] (-6,.8) to (2,.8);
\draw [black,very thick]  (-2,1.5) to (2,1);
\draw [black, very thick]  (-2,1.5) to (-6,1);
\draw [red] (-4,-1) to (-4,-.6);
\draw [red] (-4,1) to (-4,.6);
\draw [red] (0,1) to (0,.6);
\draw [red] (0,-1) to (0,-.6);
\draw [red] (-2,1.8) to (-2,1.2);
\filldraw [color = black, fill=white, very thick] (-6,0) circle (1cm)
node {$N_0^{\rm rad}$};
\filldraw [color = black, fill=white, very thick] (-2,0) circle (1cm)
node {$N_0$};
\filldraw [color = black, fill=white, very thick] (2,0) circle (1cm)
node {$N_0^{\rm rad}$};
\end{tikzpicture}
\end{gathered},\begin{gathered}\vspace{-.5cm}
\begin{tikzpicture}[scale=0.5]
\draw [black,very thick] [->] (-8,.8) to (4,.8);
\draw [black,very thick] (-8,-.8) to (-6,-.8);
\draw [black,very thick] [->] (2,-.8) to (4,-.8);
\draw [black,snake it] [->] (-6,-.8) to (2,-.8);
\draw [black,very thick]  (-2,-1.5) to (2,-1);
\draw [black, very thick]  (-2,-1.5) to (-6,-1);
\draw [red] (-4,-1) to (-4,-.6);
\draw [red] (-4,1) to (-4,.6);
\draw [red] (0,1) to (0,.6);
\draw [red] (0,-1) to (0,-.6);
\draw [red] (-2,-1.8) to (-2,-1.2);
\filldraw [color = black, fill=white, very thick] (-6,0) circle (1cm)
node {$N_0^{\rm rad}$};
\filldraw [color = black, fill=white, very thick] (-2,0) circle (1cm)
node {$N_0$};
\filldraw [color = black, fill=white, very thick] (2,0) circle (1cm)
node {$N_0^{\rm rad}$};
\end{tikzpicture}
\end{gathered}
\end{equation}
giving rise to the elementary building block

\begin{equation}
\tilde{P}_{1,2}^\mu=\langle p_1',p_2' | [\hat{N},[\hat{N},[\hat{N},\hat{P}_1^\mu]]] |p_1,p_2\rangle
\end{equation}
evaluated with two-graviton insertions. This is

\begin{align}
\tilde{P}_{1,2}^{\mu}&=G^4 \text{FT}[q^\mu \langle p_1',p_2' \vert \hat{N}_0^{\text{rad}}\hat{N}_0\hat{N}_0^{\text{rad}}\vert p_1,p_2\rangle] \cr &+G^4 \text{FT}\Big[\int \frac{d^D q_1 d^D q_2 d^D q_3}{(2\pi)^{3D-6}} (-3 q_1^{\mu}+3 q_3^{\mu})\langle p_1',p_2' \vert \hat{N}_0^{\text{rad}}\vert p_1+q_3,p_2-q_2,q_2-q_3\rangle \cr & \times  \langle p_1+q_3,q_2-q_3 \vert  \hat{N}_0 \vert p_1+q_1,q_2-q_1\rangle \delta(2p_1 \cdot q_3)\delta(-2p_2 \cdot q_2)\delta^{(+)}((q_2-q_3)^2)\cr & \times \langle p_1+q_1,p_2-q_2, q_2-q_1 \vert \hat{N}_0^{\text{rad}}\vert p_1,p_2\rangle \delta(2p_1 \cdot q_1)\delta^{(+)}((q_2-q_1)^2)\Big]+O(G^5) \cr & \equiv 6 G^4 \text{FT}[q^\mu L_2(\gamma,q^2)]+G^4 \tilde{P}_{1,2}^{\mu,(3)}+O(G^5) \cr & = 6 i G^4 p_{\infty}\frac{b^\mu}{|b|} \frac{\partial \tilde{L}_2(\gamma,J)}{\partial J}+G^4 \tilde{P}_{1,2}^{\mu,(3)}+O(G^5)
\end{align}
so that its contribution to the momentum kick becomes

\begin{equation}
{\Delta \tilde{P}_{1, \text{rad}}^{\nu,(2)}=G^4 \begin{pmatrix}
    u_1^\mu & u_2^\mu & {b^\mu\over |b|}
  \end{pmatrix}\begin{pmatrix}
0 \\
\frac{i}{6}\tilde{P}_{1,2}^{u_2,(3)}\\
-p_{\infty}\frac{\partial \tilde{L}_2(\gamma,J)}{\partial J}+ \frac{i}{6}\tilde{P}_{1,2}^{b,(3)}
\end{pmatrix}+\mathcal O(G^5)}
\end{equation}
Combining all pieces, the full fourth-order momentum kick is thus given by

\begin{equation}
\Delta \tilde{P}_{1}^{\nu,4PM}= G^4 \begin{pmatrix}
    u_1^\mu & u_2^\mu & {b^\mu\over |b|}
  \end{pmatrix}\begin{pmatrix}
p_\infty \big(- \frac{ (\chi_{\text{cons}}^{(0)})^4}{24}+\frac{(\chi_{\text{cons}}^{(1)})^2}{2}+ \chi_{\text{cons}}^{(0)}\chi_{\text{cons}}^{(2)}\big) \\
p_\infty \big( \frac{ (\chi_{\text{cons}}^{(0)})^4}{24}-\frac{(\chi_{\text{cons}}^{(1)})^2}{2}- \chi_{\text{cons}}^{(0)}\chi_{\text{cons}}^{(2)}\big)-\frac{\tilde{P}_{1,1}^{u_2,(3)}}{2}+\frac{i}{6}\tilde{P}_{1,2}^{u_2,(3)}\\
p_\infty \big(\frac{(\chi_{\text{cons}}^{(0)})^2\chi_{\text{cons}}^{(1)}}{2}-\chi_{\text{cons}}^{(3)}-\frac{\partial \tilde{L}_2(\gamma,J)}{\partial J}\big)+\frac{\mathcal E_1 \chi_{\text{cons}}^{(0)}\tilde{P}_{1,1}^{u_2,(2)}}{2} -\frac{\tilde{P}_{1,1}^{b,(3)}}{2} 
+ \frac{i}{6}\tilde{P}_{1,2}^{b,(3)} \label{MomentumKick}
\end{pmatrix}
\end{equation}
We note the partial recycling of lower-order terms here, a feature that generalizes to higher orders as well.

\section{Details on the 4PM calculation}\label{sec:4PMangle}

\subsection{The construction of the integrands}\label{sec:4PMintegrand}

To perform the full explicit computation of the momentum, we need to compute only three integrands giving $\tilde{N}^{(3)}$, $\tilde{P}_{1,1}^{\mu,(3)}$ and $\tilde{P}_{1,2}^{\mu,(3)}$. The three integrands can be represented as

\begin{align}
& \begin{gathered}
\begin{tikzpicture}[scale=0.5]
\draw [black,very thick] [->] (-2,-.8) to (2,-.8);
\draw [black,very thick] [->] (-2,.8) to (2,.8);
\filldraw [color = black, fill=white, very thick] (0,0) circle (1cm)
node {$N_3$};
\end{tikzpicture}
\end{gathered}  \\ -(q^\mu+2q_1^\mu)  \begin{gathered}
\begin{tikzpicture}[scale=0.5]
\draw [black,very thick] [->] (-8,-.8) to (0,-.8);
\draw [black,very thick] [->] (-8,.8) to (0,.8);
\draw [black,snake it]  (-5,0) to (-2,0);
\draw [red] (-4,-1) to (-4,-.6);
\draw [red] (-4,1) to (-4,.6);
\draw [red] (-4,.3) to (-4,-.3);
\filldraw [color = black, fill=white, very thick] (-6,0) circle (1cm)
node {$N_0^{\rm rad}$};
\filldraw [color = black, fill=white, very thick] (-2,0) circle (1cm)
node {$N_1^{\rm rad}$};
\end{tikzpicture}
\end{gathered} & -(q^\mu+2q_1^\mu)  \begin{gathered}
\begin{tikzpicture}[scale=0.5]
\draw [black,very thick] [->] (-8,-.8) to (0,-.8);
\draw [black,very thick] [->] (-8,.8) to (0,.8);
\draw [black,snake it]  (-5,0) to (-2,0);
\draw [red] (-4,-1) to (-4,-.6);
\draw [red] (-4,1) to (-4,.6);
\draw [red] (-4,.3) to (-4,-.3);
\filldraw [color = black, fill=white, very thick] (-6,0) circle (1cm)
node {$N_1^{\rm rad}$};
\filldraw [color = black, fill=white, very thick] (-2,0) circle (1cm)
node {$N_0^{\rm rad}$};
\end{tikzpicture}
\end{gathered}  \\  3(q_3^\mu-q_1^\mu) &\begin{gathered}\vspace{.3cm}
\begin{tikzpicture}[scale=0.5]
\draw [black,very thick] [->] (-8,-.8) to (4,-.8);
\draw [black,very thick] (-8,.8) to (-6,.8);
\draw [black,very thick] [->] (2,.8) to (4,.8);
\draw [black,snake it] [->] (-6,.8) to (2,.8);
\draw [black,very thick]  (-2,1.5) to (2,1);
\draw [black, very thick]  (-2,1.5) to (-6,1);
\draw [red] (-4,-1) to (-4,-.6);
\draw [red] (-4,1) to (-4,.6);
\draw [red] (0,1) to (0,.6);
\draw [red] (0,-1) to (0,-.6);
\draw [red] (-2,1.8) to (-2,1.2);
\filldraw [color = black, fill=white, very thick] (-6,0) circle (1cm)
node {$N_0^{\rm rad}$};
\filldraw [color = black, fill=white, very thick] (-2,0) circle (1cm)
node {$N_0$};
\filldraw [color = black, fill=white, very thick] (2,0) circle (1cm)
node {$N_0^{\rm rad}$};
\end{tikzpicture}
\end{gathered}
\end{align}
We compute these from generalized unitarity and velocity cuts, selecting topologies that both have three velocity cuts and respect the conditions on the on-shell gravitons when imposed by the topology.

\subsection{The integration basis}\label{sec:4PMintegrals}

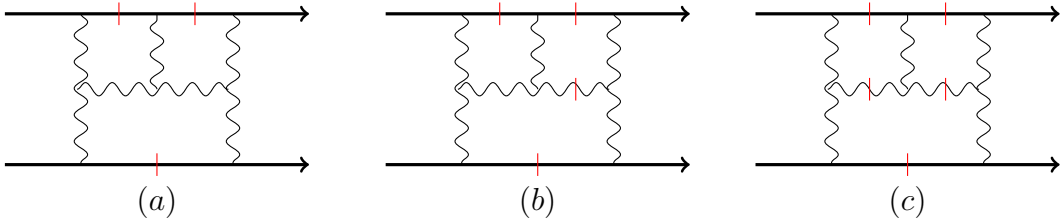
\begin{figure}[h]
  \centering
 \begin{tikzpicture}[scale=0.5]
\draw [black,very thick] [->] (-4,2) to (4,2);
\draw [black,very thick] [->] (-4,-2) to (4,-2);
\draw [black,snake it]  (-2,2) to (-2,-2);
\draw [black,snake it]  (2,2) to (2,-2);
\draw [black,snake it]  (-2.1,0) to (1.88,0);
\draw [black,snake it]  (0,0) to (0,2);
\draw [red] (-1,2.3) to (-1,1.7);
\draw [red] (1,2.3) to (1,1.7);
\draw [red] (0,-2.3) to (0,-1.7);
\draw (0,-3) node {$(a)$};
\end{tikzpicture} \qquad  \begin{tikzpicture}[scale=0.5]
\draw [black,very thick] [->] (-4,2) to (4,2);
\draw [black,very thick] [->] (-4,-2) to (4,-2);
\draw [black,snake it]  (-2,2) to (-2,-2);
\draw [black,snake it]  (2,2) to (2,-2);
\draw [black,snake it]  (-2.1,0) to (1.88,0);
\draw [black,snake it]  (0,0) to (0,2);
\draw [red] (-1,2.3) to (-1,1.7);
\draw [red] (1,2.3) to (1,1.7);
\draw [red] (0,-2.3) to (0,-1.7);
\draw [red] (1,.3) to (1,-0.3);
\draw (0,-3) node {$(b)$};
\end{tikzpicture}\qquad
 \begin{tikzpicture}[scale=0.5]
\draw [black,very thick] [->] (-4,2) to (4,2);
\draw [black,very thick] [->] (-4,-2) to (4,-2);
\draw [black,snake it]  (-2,2) to (-2,-2);
\draw [black,snake it]  (2,2) to (2,-2);
\draw [black,snake it]  (-2.1,0) to (1.88,0);
\draw [black,snake it]  (0,0) to (0,2);
\draw [red] (-1,2.3) to (-1,1.7);
\draw [red] (1,2.3) to (1,1.7);
\draw [red] (0,-2.3) to (0,-1.7);
\draw [red] (1,.3) to (1,-0.3);
\draw [red] (-1,.3) to (-1,-0.3);
\draw (0,-3) node {$(c)$};
\end{tikzpicture}
 \caption{Master integrals~\eqref{e:MasterDef1} in $(a)$
  and~\eqref{e:MasterDef2}  in $(b)$.}\label{fig:Masters}  
\end{figure} 

At fourth Post-Minkowskian order the computation of the
momentum kick is expanded on two sets of master
integrals. A first family of master integrals has
delta-function constraints on the massive legs and one graviton propagator as depicted
in fig.~\ref{fig:Masters}(a)

\begin{equation}\label{e:MasterDef2}
  \mathcal J\left(\{n_j\},\{\pm,\pm,\pm\};\gamma,\epsilon\right)=
\!\!\!  \int \frac{
    \delta(2v_1\cdot \ell_1) 
    \delta(2v_1 \cdot (\ell_1+\ell_2+\ell_3)) 
    \delta(2v_2 \cdot  (\ell_1+\ell_2))\delta(\ell_2^2)}{\prod_{i=1}^{12}D_i^{n_i}} \prod_{r=1}^3 { d^{4-2\epsilon}\ell_i\over (2\pi)^{3-2\epsilon}}.
\end{equation}
where we have defined  the propagators
\begin{align}\label{e:PropDefNew}
    D_1&=\ell_1^2,  \qquad D_2=\ell_2^2, \qquad D_3=\ell_3^2,\cr
    D_4&=(\ell_1+\ell_2)^2, \qquad  D_{5}=(\ell_2+\ell_3)^2 , \qquad  D_6=(\ell_1+\ell_2+\ell_3)^2,\cr
   D_7&=(\ell_1+\hat q)^2, D_8=(\ell_1+\ell_2+\hat q)^2,  
        D_8=(\ell_1+\ell_2+\ell_3+\hat q)^2 ,\\
\nonumber
     D_9^\pm&=\pm 2 v_1 \cdot
                                            (\ell_1+\ell_2)+i\varepsilon,
                                                       D_{10}^\pm=\pm 2v_2 \cdot \ell_1+i\varepsilon ,
  D_{11}^\pm=\pm 2v_2 \cdot (\ell_1+\ell_2+\ell_3)+i\varepsilon, 
\end{align}
with  $\hat q^2=-1$, and 
$v_i=p_i/m_i$ such that $v_i^2=1$ and $v_i\cdot q=0$ for $i=1,2$, $\gamma=v_1\cdot
v_2$.
Tensorial reductions are conveniently performed using {\tt
  LiteRed}~\cite{Lee:2012cn,Lee:2013mka} which has by default the Feynman
$+i\varepsilon$ prescription. We find that for the set of master
integrals in~\eqref{e:MasterDef2} the basis needed for longitudinal pieces has dimension 54, and the one for the transverse pieces has the same dimension.   
These master integrals have a delta-function for one of the graviton
propagators as required in the one-graviton radiative sector analyzed
in section~\ref{sec:radiation}. This delta-function breaks the symmetry between $l_2$ and $l_3$ compared to the other basis.

In the conservative sector of section~\ref{sec:conservative} it is
enough to use the smaller
set of master integrals represented in
figure~\ref{fig:Masters}(b) given by 
\begin{equation}\label{e:MasterDef1}
  \mathcal I\left(\{n_j\},\{\pm,\pm,\pm\};\gamma,\epsilon\right)=
  \int \frac{
     \delta(2v_1\cdot \ell_1) 
    \delta(2v_1 \cdot (\ell_1+\ell_2+\ell_3)) 
    \delta(2v_2 \cdot  (\ell_1+\ell_2))}{\prod_{i=1}^{12}D_i^{n_i}}\prod_{r=1}^3 { d^{4-2\epsilon}\ell_i\over (2\pi)^{3-2\epsilon}}.
\end{equation}
 The tensorial reduction gives a basis of dimension 40. 
 This basis are also sufficient to compute the second radiative term $\tilde{P}_{1,2}^{\mu,(3)}$, which differs only by the boundary conditions we impose in the static $\gamma=1$ limit.

 The world-line computation of~\cite{Dlapa:2023hsl} uses master integrals with
 delta-function velocity cuts on three massive propagators at fourth
 Post-Minkowskian order. But they have either Feynman or retarded or
 advanced propagators and in the end use a total of 576 master integrals.
 Converting the retarded (respectively advanced) propagator to a
 Feynman propagator using
 \begin{equation}
   {i\over    (\ell_0\pm i\epsilon)^2-\vec \ell^2}= {i\over
     \ell_0^2-\vec \ell^2+i\epsilon}\mp \pi \delta(\ell_0) \theta(\mp\ell_0)
 \end{equation}
 allows to expand the master integrals used in~\cite{Dlapa:2023hsl} on the
basis of master integrals in~\eqref{e:MasterDef2}.

As in ref.~\cite{Bjerrum-Bohr:2022ows} we compute the integrals by solving three differential systems of sizes $40\times 40$, $54\times54$ and $54 \times 54$ respectively. There are three regions of integration, potential-potential (PP), potential-radiation (PR) and radiation-radiation (RR). We expand all master integrals in each of these regions, which gives boundary data to solve and check the solution of the differential systems.
In the end, each master integral can be expanded on 9 independent static master integrals (6 for the transverse pieces, 3 for the longitudinal contributions) as

\begin{equation}
\mathcal I^{\perp}(\gamma)=\sum_{j=1}^{3} c_{PP,\perp}^j(\gamma) I_{PP,\perp}^j +(4(\gamma^2-1))^{-\epsilon} \sum_{j=1}^{2} c_{PR,\perp}^j(\gamma) I_{PR,\perp}^j+(4(\gamma^2-1))^{-2\epsilon}c_{RR,\perp}(\gamma) I_{RR,\perp}
\end{equation}
and

\begin{equation}
\mathcal I^{\parallel}(\gamma)=(4(\gamma^2-1))^{-\epsilon} \sum_{j=1}^{2} c_{PR,\parallel}^j(\gamma) I_{PR,\parallel}^j+(4(\gamma^2-1))^{-2\epsilon}c_{RR,\parallel}(\gamma) I_{RR,\parallel}
\end{equation}
The final step is then to compute each static master integral with the correct constraint on its graviton propagator (Feynman propagator or delta-function) according to the integrand it contributes to.

\subsection{The final result for the 4PM momentum kick}

\subsubsection{The $N$-matrix elements}

For the so-called conservative part (the $N$-matrix elements), we first recall the results up to 3PM order,

\begin{equation}
\tilde{N}^{(0)}=\frac{Gm_1 m_2(2\gamma^2-1)}{\sqrt{\gamma^2-1}} \Gamma(-\epsilon)J^{2\epsilon}
\end{equation}

\begin{equation}
\tilde{N}^{(1)}=\frac{3\pi G^2m_1^2 m_2^2(m_1+m_2)(5\gamma^2-1)}{4\sqrt{s}}\frac{1}{J}
\end{equation}

\begin{multline}
\tilde{N}^{(2)}=\frac{G^3 m_1^3 m_2^3 \sqrt{\gamma^2-1}}{s}\Bigg(\frac{s(64 \gamma^6 - 120 \gamma^4 + 60 \gamma^2-5)}{3(\gamma^2-1)^2}-\frac{4}{3} m_1 m_2 \gamma (14 \gamma^2+25)\\+\frac{4 m_1 m_2(3+12 \gamma^2-4 \gamma^4) \arccosh(\gamma)}{\sqrt{\gamma^2-1}}\\+\frac{2m_1 m_2(2\gamma^2-1)^2}{\sqrt{\gamma^2-1}}\Big(\frac{8-5\gamma^2}{3(\gamma^2-1)}+\frac{\gamma(-3+2\gamma^2) \arccosh(\gamma)}{(\gamma^2-1)^{\frac{3}{2}}} \Big) \Bigg)\frac{1}{J^2}
\end{multline}
Almost all of the 4PM part of $\tilde{N}$ was already computed in ref.~\cite{Bjerrum-Bohr:2022ows}, except for one term which we correct here.
The velocity cuts automatically eliminate super-classical terms, so that the generalized unitarity integrand arises directly from $\langle p_1',p_2'|   {\hat{T}}_3 |p_1,p_2\rangle+L_0$. To this we must add $L_1$ which precisely cancel the imaginary radiation pieces as at 3PM order. Note also that the real piece from $L_1$ is canceled by a similar computation as we did in section~\ref{sec:radiation}. At the end we get

\begin{equation}
\tilde{N}^{(3)}=\tilde{N}_{PP+RR}^{(3)}+\tilde{N}_{PR}^{(3)}+\tilde{L}_2
\end{equation}
with 

\begin{multline}
\tilde{N}^{(3)}_{PP+RR}=-\frac{G^4 (m_1+m_2)^3 m_1^4 m_2^4 \pi (\gamma^2-1)}{8 s^{\frac{3}{2}}}\cr\times\Big( \mathcal{M}_4^p+\nu (4 \mathcal{M}^t_4 \log(\frac{\sqrt{\gamma^2-1}}{2})+ \mathcal{M}^{\pi^2}_4+ \mathcal{M}^{\text{rem}}_4)\Big)\frac{1}{J^3}
\end{multline}

\begin{equation}
\tilde{N}^{(3)}_{PR}=\frac{G^4 (m_1+m_2)^3 m_1^4 m_2^4 \pi  (\gamma^2-1)}{8s^{\frac{3}{2}}} \Big(\frac{6 \nu(2\gamma^2-1)(5\gamma^2-1)\mathcal I(\gamma)}{\sqrt{\gamma^2-1}}\Big) \frac{1}{J^3}
\end{equation}
and 

\begin{equation}
\mathcal I(\gamma) \equiv \frac{16-10\gamma^2}{3(\gamma^2-1)}+\frac{2\gamma(-3+2\gamma^2) \arccosh(\gamma)}{\gamma^2-1} 
\end{equation}
where for convenience of the reader we have separated the pieces in terms of regions of integration (potential P and radiation R) and used the same notation as in ref.~\cite{Bjerrum-Bohr:2022ows}. Note that, as already observed in a different context in ref.~\cite{Georgoudis:2023lgf}, the $L_2$ Compton-like term that we have in the conservative piece will exactly cancel the one in the second radiative piece.

\subsubsection{The first radiation piece}

At 3PM order the value of the coefficient of the first radiation piece can be extracted from ref.~\cite{Herrmann:2021tct} 

\begin{equation}
\tilde{P}_{1,1}^{u_2,(2)}=\frac{2 m_1^2 m_2^3 p_{\infty}^2}{J^3} \mathcal E(\gamma)
\end{equation}
with 

\begin{multline}
\frac{\mathcal E(\gamma)}{\pi}\equiv \frac{1151 - 3336 \gamma + 3148 \gamma^2 - 912 \gamma^3 + 339 \gamma^4 - 
  552 \gamma^5 + 210 \gamma^6}{48 (\gamma^2-1)^{\frac{3}{2}}}\cr+ \frac{\gamma (-3 + 2 \gamma^2) (11 - 30 \gamma^2 + 35 \gamma^4)}{ 16 (\gamma^2-1)^2} \arccosh(\gamma)\cr- \frac{-5 + 76 \gamma - 150 \gamma^2 + 60 \gamma^3 + 35 \gamma^4}{8 \sqrt{\gamma^2-1}} \log \Big(\frac{1 + \gamma}{2}\Big)
\end{multline}
while at 4PM order we have performed the computation and find for the longitudinal part

\begin{align}
\tilde{P}_{1,1}^{u_2,(3)}&=\frac{2 m_1^2 m_2^3 p_{\infty}^3}{J^4} \Bigg(\frac{(m_1 g[1]+m_2 h[1])\pi^2}{192 (\gamma^2-1)^2}+ \frac{m_1 g[2]+m_2 h[2]}{705600 \gamma^8 (\gamma^2-1)^{\frac{5}{2}}} \cr & +\Big(\frac{m_1g[3]+m_2 h[3]}{6720 \gamma^9(\gamma^2-1)^3}+\frac{(m_1g[4]+m_2h[4]) \log(2)}{8(\gamma^2-1)^2}\Big)\arccosh(\gamma)\cr &+\Big(\frac{m_1 g[5]+m_2 h[5]}{(\gamma^2-1)^{\frac{7}{2}}}+\frac{m_1g[6]+m_2 h[6] }{(\gamma^2-1)^2}\Big)\arccosh^2(\gamma)+\frac{m_1g[7]+m_2 h[7]}{8(\gamma^2-1)^2}\arccosh(\gamma) \log(\gamma)\cr &+\frac{m_1g[8]+m_2 h[8]}{8(\gamma^2-1)^2}\Big(\arccosh(\gamma) \log(\frac{1+\gamma}{2})-2 \Li_2\Big(-\gamma+\sqrt{\gamma^2-1}\Big)\Big) \cr & +\frac{m_1g[9]+m_2 h[9]}{32(\gamma^2-1)^2} \Big(\Li_2\Big(\frac{\gamma-1}{\gamma+1}\Big)-4\Li_2\Big(\sqrt{\frac{\gamma-1}{\gamma+1}}\Big)\Big)\cr &-\frac{m_1g[10]+m_2 h[10]}{16 (\gamma^2-1)^2}\Li_2\Big(-(\gamma-\sqrt{\gamma^2-1})^2\Big)\Bigg)
\end{align}
with

\begin{align}
g[1]&=\gamma (-1485 + 4993 \gamma^2 - 3195 \gamma^4 + 
   1575 \gamma^6)\cr
g[2]&=385875 - 1837500 \gamma^2 + 7188300 \gamma^4 - 
  21241500 \gamma^6 + 767410066 \gamma^8 \cr &+ 3966858415 \gamma^{10} - 
  3429240286 \gamma^{12} - 791542442 \gamma^{14} + 393897472 \gamma^{16}\cr
g[3]&=3675 - 19950 \gamma^2 + 79800 \gamma^4 - 246540 \gamma^6 + 
 222810 \gamma^8 - 25426269 \gamma^{10} \cr &- 37185456 \gamma^{12} + 
 46406238 \gamma^{14} + 2662204 \gamma^{16} - 3592192 \gamma^{18} \cr
 g[4]&=1263 - 3883 \gamma^2 + 1065 \gamma^4 - 525 \gamma^6 \cr
 g[5]&=32 \gamma^2 (60 + 35 \gamma^2 - 59 \gamma^4 + 4 \gamma^8)\cr
 g[6]&=8 \gamma (-9 + 26 \gamma^2) \cr
 g[7]&=\gamma (1041 - 2773 \gamma^2 - 1065 \gamma^4 + 525 \gamma^6) \cr
 g[8]&=3 (37 \gamma - 185 \gamma^3 + 355 \gamma^5 - 175 \gamma^7)  \cr
 g[9]&=6 (6 - 37 \gamma - 66 \gamma^2 + 185 \gamma^3 + 210 \gamma^4 - 
   355 \gamma^5 - 150 \gamma^6 + 175 \gamma^7) \cr
 g[10]&=\gamma (1041 - 2773 \gamma^2 - 1065 \gamma^4 + 525 \gamma^6)
 \end{align}
 
\begin{align}
h[1]&=2 (2075 + 17367 \gamma^2 + 5553 \gamma^4 - 6819 \gamma^6) \cr
h[2]&=490 \gamma (1575 - 8250 \gamma^2 + 35710 \gamma^4 - 142640 \gamma^6 - 
   5560073 \gamma^8 - 417302 \gamma^{10} + 4034092 \gamma^{12}\cr & - 
   587336 \gamma^{14} + 6144 \gamma^{16}) \cr
h[3] &=14 \gamma (525 - 3100 \gamma^2 + 13690 \gamma^4 - 55260 \gamma^6 + 
   816595 \gamma^8 + 3752006 \gamma^{10}\cr & - 1978290 \gamma^{12} - 
   1029342 \gamma^{14} + 213480 \gamma^{16} + 24576 \gamma^{18}) \cr
h[4]&=-2 (2057 + 15261 \gamma^2 + 3387 \gamma^4 - 4321 \gamma^6) \cr
h[5]&=-32 \gamma (-3 + 2 \gamma^2) (-8 - 51 \gamma^2 - 6 \gamma^4 + 8 \gamma^6)  \cr
h[6]&=16 (16 + 111 \gamma^2 + 18 \gamma^4 - 24 \gamma^6) \cr
h[7]&=-2 (2039 + 13155 \gamma^2 + 1221 \gamma^4 - 1823 \gamma^6) \cr
h[8]&=-2 (9 + 1053 \gamma^2 + 1083 \gamma^4 - 1249 \gamma^6)   \cr
h[9]&=6   (36 - 1209 \gamma + 4212 \gamma^2 - 6422 \gamma^3 + 4332 \gamma^4 + 
  1755 \gamma^5 - 4996 \gamma^6 + 2100 \gamma^7) \cr
h[10]&=-2 (2039 + 13155 \gamma^2 + 1221 \gamma^4 - 1823 \gamma^6)
\end{align}

For the transverse part we find

\begin{multline}
\tilde{P}_{1,1}^{b,(3)}=-\frac{2m_1^2 m_2^2 p_{\infty}^4}{J^4}\Bigg(\Big(-\frac{2 \gamma^2-1}{\gamma^2-1} \mathcal C(\gamma)+\frac{\gamma(-3+2\gamma^2)}{(\gamma^2-1)^{\frac{3}{2}}} \mathcal E(\gamma)\Big)(m_1+m_2)\cr+\frac{2\gamma^2-1}{(\gamma+1)\sqrt{\gamma^2-1}}\mathcal E(\gamma) m_1\Bigg)
\end{multline}
with

\begin{multline}
\frac{\mathcal C(\gamma)}{\pi}\equiv \frac{-237 + 386 \gamma + 111 \gamma^2 - 683 \gamma^3 + 537 \gamma^4 + 
 240 \gamma^5 - 411 \gamma^6 + 105 \gamma^7}{24 (\gamma^2-1)^2}\\-\frac{\gamma (-3 + 2 \gamma^2) (-12 + 19 \gamma + 72 \gamma^2 - 70 \gamma^3 - 60 \gamma^4 + 35 \gamma^5) }{8 (\gamma^2-1)^{\frac{5}{2}}}\arccosh(\gamma)\\+ \frac{-62 + 155 \gamma + 16 \gamma^2 - 70 \gamma^3 - 90 \gamma^4 + 
   35 \gamma^5}{4 (\gamma^2-1)}\log(\frac{1 + \gamma}{2})
\end{multline}

\subsubsection{The second radiation piece}

The contributions from the second radiation piece matches exactly the result of ref.~\cite{Dlapa:2022lmu} with

\begin{equation}
\tilde{P}_{1,3}^{b,(3)}=-\frac{6 i p_{\infty}^4}{J^4} c_{\text{1b,2rad}}^{(4) \text{diss}}
\end{equation}
and

\begin{equation}
\tilde{P}_{1,3}^{u_2,(3)}=-\frac{6 i m_2 p_{\infty}^3}{J^4} c_{\text{1} \check{u}_2\text{,2rad}}^{(4) \text{diss}}
\end{equation}
Finally, when inserting all integrals into the formula of eq. (\ref{MomentumKick}) we find complete agreement with ref.~\cite{Dlapa:2022lmu}.
This amplitude-based approach, which combines the exponential representation of the gravitational $S$-matrix
with the KMOC formalism, thus yields a result for the momentum kick that is in full agreement with the worldline calculation of ref.~\cite{Dlapa:2022lmu}.


\section{Conclusion}\label{sec:conclusion}

The exponential representation of the $S$-matrix \cite{Damgaard:2021ipf} is a natural starting point for a semi-classical analysis
of quantum field theory. Matrix elements of the $\hat N$-operator in the exponent of the $S$-matrix are by construction free of
superclassical terms and they are, therefore, at leading order
providing the classical part, followed by quantum corrections.
Using the KMOC-formalism,  we have shown how the exponential
representation of the $S$-matrix makes manifest the cancellation of
superclassical contributions in the conservative sector. One advantage of working with the $\hat N$-matrix rather than the conventional $\hat T$-matrix is indeed
that it bypasses the need to
ensure the delicate cancellation between superclassical terms of the $\hat T$-matrix. Instead, by extracting the relevant pieces
of the $\hat N$-matrix by means of velocity cuts we automatically retrieve the classical terms. Pictorially speaking,
the velocity cuts introduced in~\cite{Bjerrum-Bohr:2021din} localize the massive scattering states on classical on-shell trajectories.
As shown  in section~\ref{sec:conservative}  of the present paper, two-to-two massive matrix
elements of the $\hat N$-operator, Fourier-transformed to impact parameter space, is the
radial action of the conservative sector. This proves the conjectured relation put forward in 
ref.~\cite{Damgaard:2021ipf}.

Including gravitational radiation, the $\hat N$-operator is still a basic building block of the KMOC-formalism and as an example we have
shown how the momentum kick in the scattering of two
black holes can be compactly described by matrix elements of $\hat{N}$. We have provided the explicit formulas up to and
including fourth Post-Minkowskian order but the framework is iterative and it is straightforward to derive corresponding
expressions to arbitrarily high order in Newton's constant $G$.  As an
application we have explicitly derived the momentum kick at fourth
post-Minkowskian order. Our results are in agreement
with~\cite{Dlapa:2022lmu,Dlapa:2023hsl}. As is well known, and somewhat disturbingly, it leads to a
scattering angle that diverges at high energy if one applies the scattering angle expression of ref. \cite{Bini:2021gat}. The solution for the integrals
used here and in the references above is the one connecting smoothly to the Post-Newtonian expansion. We cannot exclude that another solution exists which
is valid at high energy only and without a smooth connection to the
Post-Newtonian limit. This possibility seems to deserve attention. Alternatively, one could consider doing a new fourth-order calculation from scratch with
massless scalars.

The resulting relationship between the KMOC-formalism and the exponential representation of the $S$-matrix is very simple and 
of a universal form involving trigonometric functions together with iterated commutators. This trigonometric structure arises from $\hat{N}$ being the 
exponential phase operator of the $S$-matrix and is thus closely linked to the Euler formula. Beyond the conservative parts, the operator identities involved
lead to additional terms but the structure of nested commutators is
responsible for the simple algebraic relations that iteratively build up observables to higher and higher orders in the gravitational coupling 
constant.

In the end, the expression for classical observables including all dissipative effects becomes remarkably simple by combining the KMOC formalism
with the exponential representation of the $S$-matrix. The full calculation reduces to 
scattering amplitude evaluations
for which modern techniques have become highly developed. There is thus no need to distinguish between different pieces or to separate the amplitude calculation 
into different types of  contributions; one must only retain all classical terms, as this provides the full classical answer.

\subsection*{Acknowledgements}
We thank Thibault Damour for comments.
P.V. would like to thank the LAPTh for the hospitality during the
completion of this work.
The work of P.H.D. was supported in part by DFF grant 0135-00089A,
the work of E.R.H. was supported by the Rozenthal Foundation and
ERC Starting Grant No. 757978 from the European Research Council, and 
the research of P.V. has received funding from the ANR grant ``SMAGP''
ANR-20-CE40-0026-01. 


\end{document}